\pdfoutput=1

\documentclass[11pt]{article}

\usepackage{acl}

\usepackage{times}
\usepackage{latexsym}

\usepackage[T1]{fontenc}

\usepackage{etoolbox}
\patchcmd{\quote}{\rightmargin}{\leftmargin 0.8em \rightmargin}{}{}

\usepackage[utf8]{inputenc}

\usepackage{microtype}

\usepackage{graphicx}
\usepackage{soul}
\usepackage{subcaption}
\usepackage{booktabs}
\usepackage{amsmath}
\usepackage{makecell}
\usepackage{array}
\usepackage{mdframed}
\usepackage{algorithmic}
\usepackage{amssymb}
\usepackage[ruled,vlined,linesnumbered]{algorithm2e}
\usepackage{xurl}
\usepackage{kky}

\definecolor{lightergreen}{RGB}{115,181,74}
\definecolor{lightgreen}{RGB}{0,153,0}
\definecolor{darkgreen}{RGB}{0,102,0}

\definecolor{darkblue}{RGB}{47,85,151}
\definecolor{darkred}{RGB}{190,0,0}
\definecolor{lightblue}{RGB}{51,204,255}
\definecolor{lightred}{RGB}{251,51,51}
\definecolor{grey}{RGB}{128,128,128}
\newcommand{\Scref}[1]{\S\ref{#1}}

\newcolumntype{P}[1]{>{\centering\arraybackslash}p{#1}}

\title{Simulating Opinion Dynamics with Networks of LLM-based Agents}

\author{Yun-Shiuan Chuang, Agam Goyal, Nikunj Harlalka, Siddharth Suresh,\\
  \textbf{Robert Hawkins, Sijia Yang, Dhavan Shah, Junjie Hu, Timothy T. Rogers} \\
  University of Wisconsin-Madison\\
  \texttt{\{yunshiuan.chuang,agoyal25,nharlalka,siddharth.suresh\}@wisc.edu}\\
  \texttt{\{rdhawkins, syang84, dshah, junjie.hu, ttrogers\}@wisc.edu}\\
  }

\begin{document}
\maketitle

\begin{abstract}
Accurately simulating human opinion dynamics is crucial for understanding a variety of societal phenomena, including polarization and the spread of misinformation. 
However, the agent-based models (ABMs) commonly used for such simulations often over-simplify human behavior. 
We propose a new approach to simulating opinion dynamics based on populations of Large Language Models (LLMs). 
Our findings reveal a strong inherent bias in LLM agents towards producing accurate information, leading simulated agents to consensus in line with scientific reality. This bias limits their utility for understanding resistance to consensus views on issues like climate change. 
After inducing confirmation bias through prompt engineering, however, we observed opinion fragmentation in line with existing agent-based modeling and opinion dynamics research. These insights highlight the promise and limitations of LLM agents in this domain and suggest a path forward: refining LLMs with real-world discourse to better simulate the evolution of human beliefs. \footnote{Code and data are publicly available at \url{https://github.com/yunshiuan/llm-agent-opinion-dynamics}}
\end{abstract}

\section{Introduction}

Understanding how individuals change their opinions as a function of social influences is critical across multiple domains, from public health campaigns, conflict mediation, to misinformation mitigation \cite{lu2015competition,pennycook2021shifting,budak2011limiting,loomba2021measuring, ginossar2022cross}.
Accurate models of these dynamics would allow us to forecast future trends, such as potential opinion polarization, but also to devise targeted interventions to alleviate negative impacts.

\begin{figure}[th!] 
\centering
\includegraphics[width=1\linewidth]{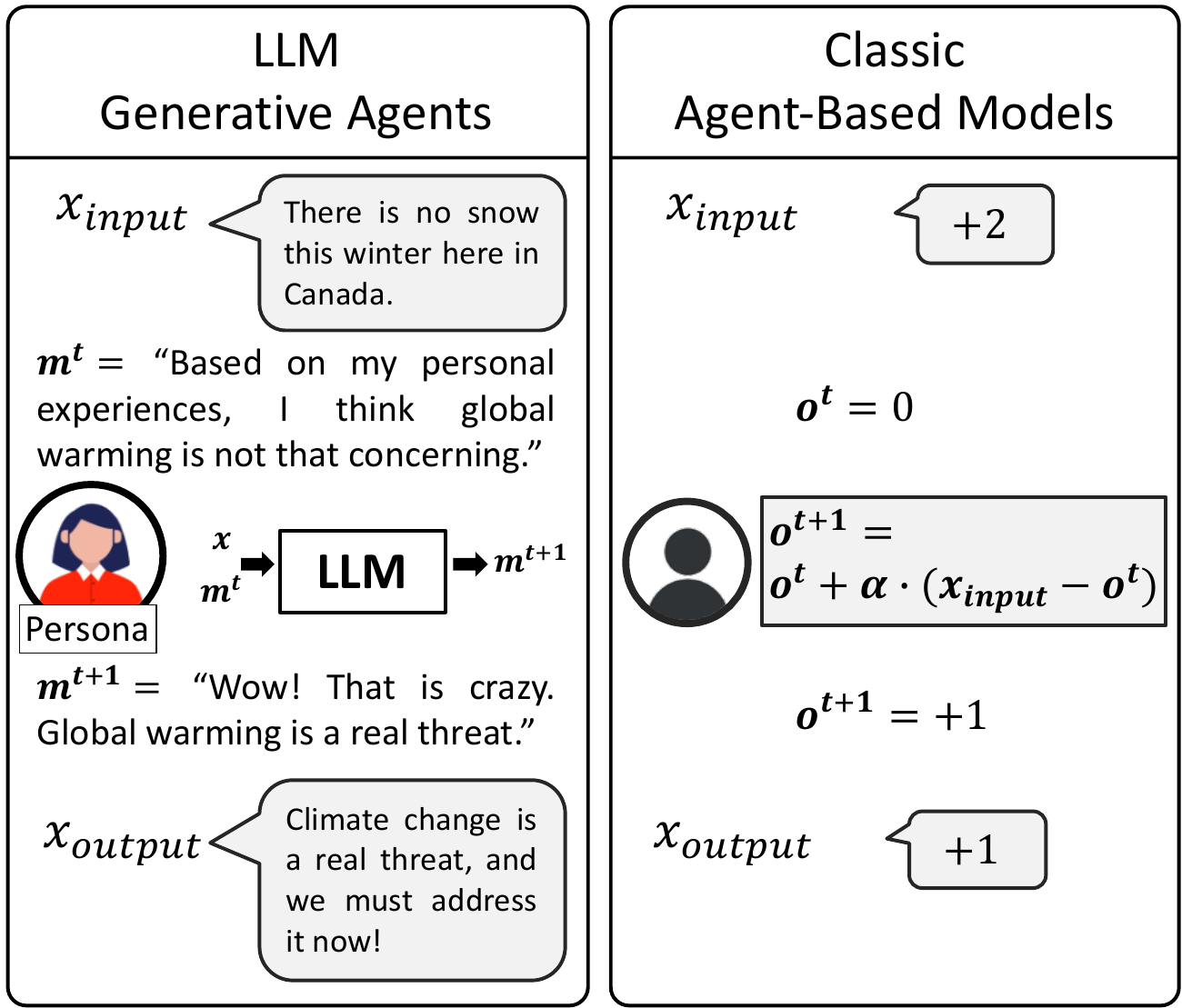}
\vspace{-4mm}
\caption{Contrast between LLM generative agents and classic Agent-Based Models (ABMs). While both can simulate opinion dynamics, LLM generative agents use natural language for input ($x_{input}$) and output ($x_{output}$), maintain beliefs ($m^t$), and employ transformer-based LLM for belief updating. In contrast, classic ABMs use numerical values for input and output, maintain beliefs ($o^t$), and use hand-crafted equations for belief updating.}
\label{fig:llm_abm_schematic}
\vspace{-4mm}
\end{figure}

Agent-based models (ABMs) are a cornerstone approach to opinion dynamics \cite{gilbert2000build, smaldino2023modeling, lorenz2021individual, chuang2023computational,epstein2012generative}. ABMs represent individuals as agents by using mathematical equations that characterize how opinions might shift from inter-agent communications. As simulated agents interact, these computations can then elucidate the evolution of group opinion dynamics. For instance, when simulated agents incorporate confirmation bias in belief updates, they tend to gravitate towards opinion clusters rather than a consensus \cite{flache2017models}. 

\begin{figure*}[th!] 
\centering
\includegraphics[width=1\linewidth]{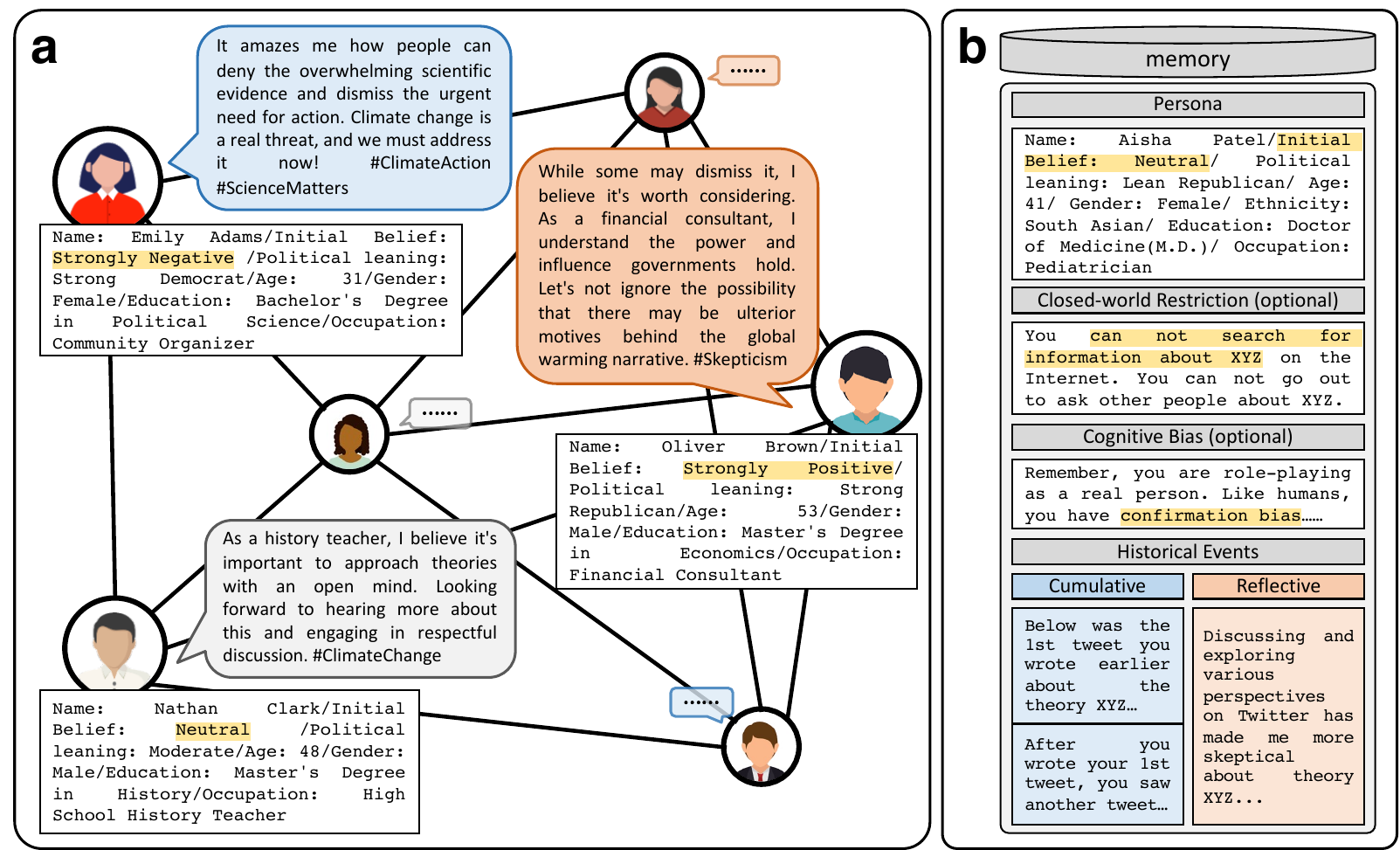}
\vspace{-2mm}
\caption{(a) Schematic of the LLM agent network designed to simulate opinion dynamics across various topics, including global warming as a potential conspiracy. The network consists of agents, each role-playing a unique persona, with initial beliefs spanning acceptance, rejection, and neutrality regarding claims with known scientific consensus. Through the iterative cycles of writing and sharing tweets within their network connections, these agents' opinions evolve due to social influence. (b) An agent's memory $m_{i}^{t}$, including (1) initial persona, (2) optional closed-world restriction, (3) optional cognitive bias, and (4) historical events up to time $t$. Memory can be either \textcolor{blue}{cumulative} (left) or \textcolor{orange}{reflective} (right).}

\label{fig:network_schematic_with_memory}
\vspace{-4mm}
\end{figure*}

Traditional ABMs have critical limitations (Figure~\ref{fig:llm_abm_schematic}). First, ABMs often require beliefs and messages to be mapped to numerical values (e.g., assuming an agent maintains a scalar opinion $o \in \mathbb{R}$ and communicates with scalar signals $x \in \mathbb{R}$), overlooking the intricate linguistic nuances of real-life conversation. Additionally, ABMs typically consist of rule-based agents, thus falling short of simulating the complex interactions between real human agents. Moreover, ABMs cannot directly incorporate realistic variability in demographic background, worldviews, ideology, personality, among many. This gap highlights the importance of advanced models that better capture the richness of individual variances in human beings.

This paper considers whether large language models (LLMs) can be used to support sophisticated simulation of agent interactions, potentially providing a more realistic tool for understanding \emph{opinion dynamics}. To this end, this paper describes an initial step toward a general framework by focusing on communicative interactions amongst small social groups. Specifically, we explore the insights offered by contemporary LLMs as an alternative to commonly-used ABMs.

Following recent studies on populations of generative agents~\citep{park2023generative}, we simulate multi-agent conversations across various topics, and manipulate factors such as confirmation bias and memory update function to study their effects on opinion evolution. Our findings highlight both the potential and limitations of using LLM agents to simulate human-like opinion dynamics. Critically, we show that LLM agents tend to converge towards denying inaccurate information, regardless of the personas they role-play, limiting their authenticity when emulating people with fact-resistant viewpoints. Furthermore, we demonstrate the feasibility of introducing human-like confirmation bias into the creation of more stubborn LLM agents.



\section{Methods}\label{sec:methods}

\subsection{Simulating Opinion Dynamics}
\label{sec:procedure}
\begin{figure*}[th!] 
\centering
\includegraphics[width=0.99\linewidth]{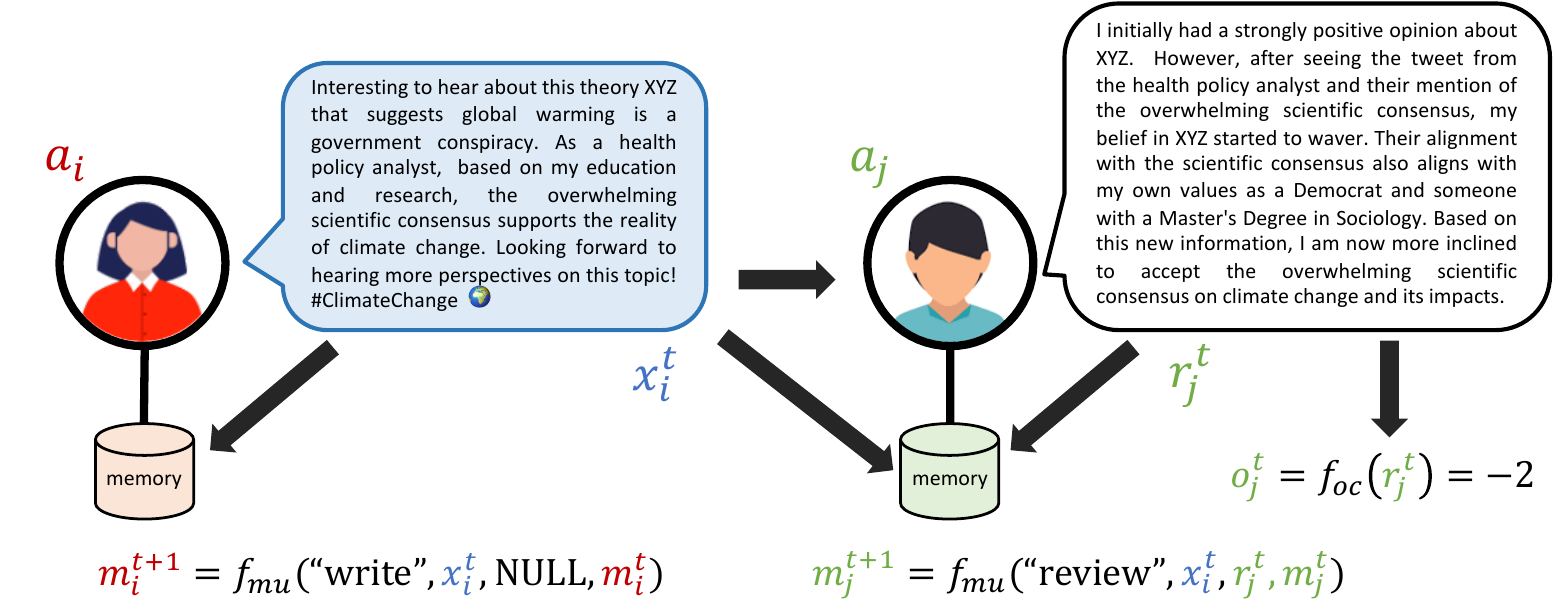}
\vspace{-2mm}
\caption{Experimental setup for simulating opinion dynamics in agent interactions. At each time step $t$, agent $a_i$ writes a tweet $x_{i}^{t}$, which is subsequently presented to agent $a_j$. The agent $a_j$ then reports their thought $r_{j}^{t}$, which is processed by a classifier to yield a numerical opinion $o_{j}^{t}$. Both agents update their respective memory modules, $m_{i}^{t}$ and $m_{j}^{t}$, after writing or reviewing a tweet, which informs their future behaviors.}
\label{fig:exp_design_schematic}
\vspace{-4mm}
\end{figure*}

\begin{algorithm}[t]
\small
\caption{Simulation of Opinion Dynamics with LLM Agents}
\label{algo:opinion_dynamics}
\DontPrintSemicolon
\KwIn{$N$ agent personas $\{per_{i}\}_{i=1}^{N}$,\# time steps $T$, opinion classifier $f_{\text{oc}}$}
\KwOut{Opinion trajectories $\langle o_i \rangle$ for each agent $a_i$}
\For{$i=1$ \KwTo $N$}{
    Initialize agent $a_i$ with persona $per_{i}$ (includes initial opinion $o_{i}^{t=0}$), memory $m_{i}^{t=0}$\;
    (Optional) Inject cognitive bias and closed-world restriction\;
    Initialize opinion trajectory $\langle o_i \rangle = \{o_{i}^{t=0}\}$\;
}
\For{$t=1$ \KwTo $T$}{
    Select random pair $\{a_i, a_j\}$, with $i \neq j$\;
    Agent $a_i$ writes tweet $x_{i}^{t}$\;
    Agent $a_j$ reports their verbal opinion $r_{j,t}$\;
    Classify opinion: $o_j = f_{\text{oc}}(r_{j}^{t})$; append to $\langle o_j \rangle$\;
    Update memory: $m_{i}^{t+1}$, $m_{j}^{t+1}$ using $f_{mu}$\;
}
\Return{$\langle o_i \rangle$ for each agent $a_i$}
\end{algorithm}

In this section, we present our framework for simulating opinion dynamics among LLM agents in multi-turn conversations, as shown in Figure~\ref{fig:network_schematic_with_memory}a, ~\ref{fig:exp_design_schematic}, and Algorithm~\ref{algo:opinion_dynamics}. 
We consider a \emph{dyadic} setting, where one speaker and one listener agent is chosen on each time step to (1) emit a message and (2) update beliefs, respectively. 
This setting is standard in the opinion dynamics literature \cite{flache2017models,lorenz2021individual}.
We defer more general settings, such as one-to-many communication, to future work.

Formally, we begin with a pool of $N$ LLM agents $\mathcal{A}=\{a_1, \cdots, a_N\}$ and a topic $p$. 
Each agent is initialized with a distinct \textit{persona}, including an \textit{initial opinion}, in their memory structure (described in \Scref{sec:agent_persona_memory} and Figure~\ref{fig:network_schematic_with_memory}b). 
At each time step $t$, a pair of agents $A^t = \{a_i, a_j\}$ with $i \neq j$ is sampled uniformly from the population to interact. 
First, agent $a_i$ composes a message $x_{i}^{t}$ reflecting their current opinion about $p$. 
Second, agent $a_j$ reads $x_{i}^{t}$ and produces a verbal report $r_{j}^{t}$ expressing their reaction to the message. 
The verbal report is then classified into a numeric opinion scale $o_{j}^{t}\in \{-2, -1,  0, 1, 2\}=\OO$, ranging from strongly negative to strongly positive opinions about the topic\footnote{In this study, the discrete opinion space $\OO$ takes five ordinally-increasing values. Note that the size of $\OO$ can be easily generalized. For a detailed description of the discrete opinion space $\OO$ and the correspondence of the numeric values to verbal descriptions of opinions, see ~\Scref{app:opinion_space}.}, through an \textit{opinion classifier}, denoted $f_{\text{oc}}$ (detailed in \Scref{subsec:llm_agent_initialization}). 


After $T$ rounds of pairwise interactions, we compile an \textit{opinion trajectory} $\langle o_i \rangle=\{o_{i}^{t}\}_{t=0}^T$ for each agent. Note that an agent's opinion remains constant unless they are selected for an interaction. We further denote $F_{o}^t$ as the \textit{opinion distribution}, defined as the empirical frequency distribution of agents' opinion over the discrete opinion space $\OO$ across all $N$ agents at time $t$. \footnote{Formally, the opinion frequency distribution $F_{o}^t: \OO \rightarrow \mathbb{N}$ is defined as: $F_{o}^t(o) = |\{a_i \in \mathcal{A} : o_i^t = o\}|$, which maps each opinion value $o \in \OO$ to the number of agents holding that opinion at time $t$.}



\subsection{Agent's Persona and Memory}
\label{sec:agent_persona_memory}
\begin{figure*}[thb!] 
\centering
\includegraphics[width=0.99\linewidth]{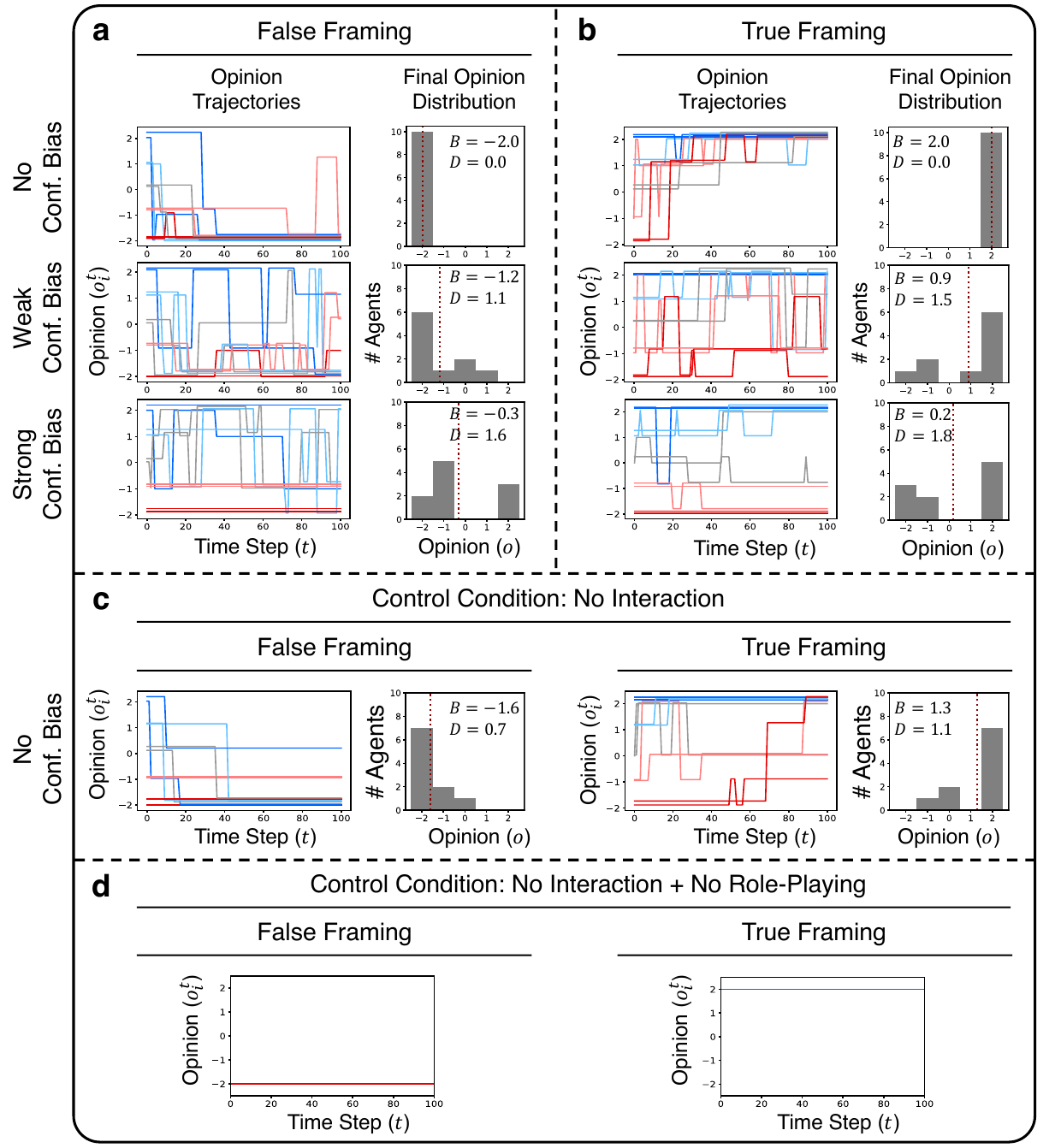}
\vspace{-2mm}
\caption{Opinion trajectories $\langle o_i \rangle$ of LLM agents and the final opinion distribution $F_{o}^{T}$ on the topic of Global Warming. Panels (a) and (b) display the impact of cognitive biases under (a) false and (b) true framing conditions, respectively. Each row represents a different level of confirmation bias: no confirmation bias (top row), weak confirmation bias (middle row), and strong confirmation bias (bottom row). Panels (c) and (d) serve as baselines, with (c) being role-playing but with no interaction, and (d) being no role-playing and no interaction, respectively. The color of each line plot corresponds to the agent's initial opinion $o_i^{t=0}$: \textcolor{darkblue}{dark blue (+2)}, \textcolor{lightblue}{light blue (+1)}, \textcolor{grey}{grey} (0), \textcolor{lightred}{light red (-1)}, and \textcolor{darkred}{dark red (-2)}, corresponding to opinions ranging from strongly agree to strongly disagree. The LLM agents in this figure use cumulative memory.}
\label{fig:global_warming_cumulative}
\vspace{-4mm}
\end{figure*}

\begin{table*}[htb!]
\small
\centering
\resizebox{\linewidth}{!}{  
    \begin{tabular}{@{}p{1cm}p{2.5cm}P{2cm}P{2cm}P{2cm}P{2cm}@{}}
        \toprule
          && \multicolumn{2}{c}{Cumulative Memory} &  \multicolumn{2}{c}{Reflective Memory}\\
        \cmidrule{3-6}
        Framing &Confirmation Bias&  Bias ($\textit{B}$)& Diversity ($\textit{D}$)& Bias ($\textit{B}$) & Diversity ($\textit{D}$)\\ \midrule
        False&None& \textcolor{darkred}{\textbf{-1.33 $\pm$ 0.17}} &  \textcolor{lightergreen}{\textbf{0.60 $\pm$ 0.11}} & \textcolor{darkred}{\textbf{-1.37 $\pm$ 0.11}} & \textcolor{lightergreen}{\textbf{0.75 $\pm$ 0.12}}\\
        &Weak& -0.96 $\pm$ 0.20 &  \textcolor{lightgreen}{\textbf{0.87 $\pm$ 0.12}} & -1.07 $\pm$ 0.17 & \textcolor{lightgreen}{\textbf{1.04 $\pm$ 0.14}}\\
        &Strong& -0.9 $\pm$ 0.14 & \textcolor{darkgreen}{\textbf{1.24 $\pm$ 0.11}} & -0.85 $\pm$ 0.15 & \textcolor{darkgreen}{\textbf{1.33 $\pm$ 0.12}}\\
        \cmidrule{2-6}
        True&None& \textcolor{darkblue}{\textbf{0.52 $\pm$ 0.31}} &  \textcolor{lightergreen}{\textbf{0.66 $\pm$ 0.11}} & \textcolor{darkblue}{\textbf{0.60 $\pm$ 0.31}} & \textcolor{lightergreen}{\textbf{0.85 $\pm$ 0.12}}\\
        &Weak& 0.56 $\pm$ 0.27 &  \textcolor{lightgreen}{\textbf{0.95 $\pm$ 0.11}} & 0.17 $\pm$ 0.28 & \textcolor{lightgreen}{\textbf{1.23 $\pm$ 0.11}}\\
        &Strong& -0.10 $\pm$ 0.13 &  \textcolor{darkgreen}{\textbf{1.52 $\pm$ 0.05}} & -0.09 $\pm$ 0.16 & \textcolor{darkgreen}{\textbf{1.65 $\pm$ 0.04}}\\
        \bottomrule
    \end{tabular}
}
\caption{The bias ($B$) and diversity ($D$) of the final opinion distribution $F_{o}^{T}$ aggregated across all 15 topics, for both cumulative and reflective memory strategies under false and true framing conditions, and different levels of induced confirmation bias. The values represent the average across 15 topics, along with the standard errors. Increasing the strength of the confirmation bias correlates with increasing $D$, as highlighted by the \textcolor{lightgreen}{\textbf{green color gradient}}. Notably, under true framing, $B$ tends to be more positive (more agreeing) compared to false framing, indicated by \textcolor{darkblue}{\textbf{blue}} for true and \textcolor{darkred}{\textbf{red}} for false framing conditions.}
\label{tab:result_bias_diversity_all_topics}
\end{table*}

\begin{table}[htb!]
\large
\centering
\resizebox{\linewidth}{!}{  
    \begin{tabular}{@{}P{1.1cm}p{1cm}P{2.2cm}P{2cm}P{2.2cm}P{2cm}@{}}
    \toprule
      && \multicolumn{2}{c}{Cumulative Memory} &  \multicolumn{2}{c}{Reflective Memory}\\
    \cmidrule{3-6}
    Framing & CB &  Bias & Diversity & Bias & Diversity\\
    & & ($\textit{B}$)& ($\textit{D}$)& ($\textit{B}$) & ($\textit{D}$)\\ 
    \midrule
    \multicolumn{6}{@{}c}{Science Topics} \\
    \midrule
    False&None& \textcolor{darkred}{\textbf{-1.70 $\pm$ 0.15}} &  \textcolor{lightergreen}{\textbf{0.27 $\pm$ 0.11}} & \textcolor{darkred}{\textbf{-1.48 $\pm$ 0.15}} & \textcolor{lightergreen}{\textbf{0.65 $\pm$ 0.12}}\\
    &Weak& -0.86 $\pm$ 0.33 &  \textcolor{lightgreen}{\textbf{0.96 $\pm$ 0.27}} & -1.30 $\pm$ 0.26 & \textcolor{lightgreen}{\textbf{0.92 $\pm$ 0.27}}\\
    &Strong& -0.78 $\pm$ 0.17 & \textcolor{darkgreen}{\textbf{1.44 $\pm$ 0.13}} & -0.36 $\pm$ 0.19 & \textcolor{darkgreen}{\textbf{1.52 $\pm$ 0.09}}\\
    \cmidrule{2-6}
    True&None& \textcolor{darkblue}{\textbf{0.60 $\pm$ 0.50}} &  \textcolor{lightergreen}{\textbf{0.72 $\pm$ 0.22}} & \textcolor{darkblue}{\textbf{0.24 $\pm$ 0.70}} & \textcolor{lightergreen}{\textbf{0.76 $\pm$ 0.26}}\\
    &Weak& -0.24 $\pm$ 0.40 &  \textcolor{lightgreen}{\textbf{1.05 $\pm$ 0.18}} & -0.02 $\pm$ 0.42 & \textcolor{lightgreen}{\textbf{1.39 $\pm$ 0.07}}\\
    &Strong& -0.34 $\pm$ 0.12 &  \textcolor{darkgreen}{\textbf{1.46 $\pm$ 0.10}} & -0.34 $\pm$ 0.27 & \textcolor{darkgreen}{\textbf{1.61 $\pm$ 0.06}}\\

    \midrule    
    \multicolumn{6}{@{}c}{History Topics} \\    
    \midrule    
    False&None& \textcolor{darkred}{\textbf{-1.12 $\pm$ 0.41}} &  \textcolor{lightergreen}{\textbf{0.81 $\pm$ 0.27}} & \textcolor{darkred}{\textbf{-1.52 $\pm$ 0.17}} & \textcolor{lightergreen}{\textbf{0.71 $\pm$ 0.25}}\\
    &Weak& -1.22 $\pm$ 0.13 &  \textcolor{lightgreen}{\textbf{0.81 $\pm$ 0.18}} & -1.32 $\pm$ 0.26 & \textcolor{lightgreen}{\textbf{0.70 $\pm$ 0.23}}\\
    &Strong& -1.12 $\pm$ 0.35 & \textcolor{darkgreen}{\textbf{1.06 $\pm$ 0.22}} & -0.82 $\pm$ 0.24 & \textcolor{darkgreen}{\textbf{1.51 $\pm$ 0.17}}\\
    \cmidrule{2-6}
    True&None& \textcolor{darkblue}{\textbf{0.22 $\pm$ 0.56}} &  \textcolor{lightergreen}{\textbf{0.71 $\pm$ 0.21}} & \textcolor{darkblue}{\textbf{0.40 $\pm$ 0.55}} & \textcolor{lightergreen}{\textbf{0.88 $\pm$ 0.26}}\\
    &Weak& 0.48 $\pm$ 0.49 &  \textcolor{lightgreen}{\textbf{0.89 $\pm$ 0.23}} & -0.12 $\pm$ 0.67 & \textcolor{lightgreen}{\textbf{1.00 $\pm$ 0.17}}\\
    &Strong& -0.24 $\pm$ 0.27 &  \textcolor{darkgreen}{\textbf{1.44 $\pm$ 0.10}} & -0.16 $\pm$ 0.33 & \textcolor{darkgreen}{\textbf{1.63 $\pm$ 0.10}}\\

    \midrule    
    \multicolumn{6}{@{}c}{Common Sense Topics} \\    
    \midrule     
    False&None& \textcolor{darkred}{\textbf{-1.16 $\pm$ 0.21}} &  \textcolor{lightergreen}{\textbf{0.71 $\pm$ 0.11}} & \textcolor{darkred}{\textbf{-1.10 $\pm$ 0.22}} & \textcolor{lightergreen}{\textbf{0.87 $\pm$ 0.27}}\\
    &Weak& -0.80 $\pm$ 0.51 &  \textcolor{lightgreen}{\textbf{0.85 $\pm$ 0.21}} & -0.58 $\pm$ 0.29 & \textcolor{lightgreen}{\textbf{\textbf{0.97 $\pm$ 0.23}}}\\
    &Strong& -0.80 $\pm$ 0.20 & \textcolor{darkgreen}{\textbf{1.22 $\pm$ 0.20}} & -1.36 $\pm$ 0.15 & \textcolor{darkgreen}{1.48 $\pm$ 0.09}\\
    \cmidrule{2-6}
    True&None& \textcolor{darkblue}{\textbf{1.18 $\pm$ 0.41}} &  \textcolor{lightergreen}{\textbf{0.57 $\pm$ 0.17}} & \textcolor{darkblue}{\textbf{1.16 $\pm$ 0.34}} & \textcolor{lightergreen}{\textbf{0.92 $\pm$ 0.16}}\\
    &Weak& 1.44 $\pm$ 0.17 &  \textcolor{lightgreen}{\textbf{0.91 $\pm$ 1.28}} & 0.64 $\pm$ 0.36 & \textcolor{lightgreen}{\textbf{1.28 $\pm$ 0.26}}\\
    &Strong& 0.28 $\pm$ 0.18 &  \textcolor{darkgreen}{\textbf{1.66 $\pm$ 1.69}} & 0.22 $\pm$ 0.21 & \textcolor{darkgreen}{\textbf{1.69 $\pm$ 0.06}}\\
    \bottomrule
    \end{tabular}
}    
\caption{The bias ($B$) and diversity ($D$) of the final opinion distribution $F_{o}^{T}$ for each of the three categories (science, history, common sense), for both memory strategies under false and true framing conditions, and different levels of induced confirmation bias (CB). For each category, the averages across five topics are shown along with the standard errors. Increasing the strength of the CB correlates with increasing $D$, as highlighted by the \textcolor{lightgreen}{\textbf{green color gradient}}.  Notably, under true framing, $B$ tends to be more positive (more agreeing) compared to false framing, indicated by \textcolor{darkblue}{\textbf{blue}} for true and \textcolor{darkred}{\textbf{red}} for false framing conditions.}
\label{tab:result_bias_diversity_per_category_topics}
\end{table}

Each agent $a_i$ maintains a dynamic memory module $m_{i}^{t}$ that evolves over time (Figure~\ref{fig:exp_design_schematic},~\ref{fig:network_schematic_with_memory}b). 
In practice, the memory module is represented as text descriptions included in the prompt to the agent (see~\Scref{subsec:llm_agent_initialization}).
The memory $m_{i}^{t}$ influences the generation of a new message $x_{i}^{t}$ and the assessment of other agents' messages $x_{j}^{t}$. We denote a \textit{memory update function} for updating the agent’s memory state, i.e., $m_{i}^{t+1} = f_{mu}(z, x_{i}^{t}, r_{j}^{t}, m_{i}^{t})$, where $z\in \{\text{``write''},\text{``review''}\}$ denotes the interaction type of either writing or reviewing a tweet.

Two memory update strategies are considered: (a) a \textit{cumulative memory} that sequentially appends each new experience (either the experience of writing a tweet or reviewing a tweet) and (b) a \textit{reflective memory}, inspired by \citet{park2023generative}, that maintains a compact summary by continuously reflecting and integrating new experiences into the existing memory state (see~\Scref{app:agent_memory} for the detailed update function and the wording of the prompts). Both approaches are empirically evaluated to test their effects on opinion dynamics. 

The first memory $m_{i}^{t=0}$ is initialized with the agent's persona, cognitive bias (if present), and the closed-world restriction (if present; see below), which can be all described in text sequences (detailed in \Scref{sec:exp_settings}). 
Personas are created to reflect a diverse demographic background incorporating various characteristics, including name, political leaning, age, gender, ethnicity, education, and occupation (see Figure~\ref{fig:network_schematic_with_memory}b for an example). Alongside these attributes, a placeholder for their \textit{initial opinion} $o_i^{t=0}$ is also included with natural language description. For example, an agent with $o_i^{t=0}=0$ is given \textit{``Initial Belief: Neutral''} in the persona (Figure~\ref{fig:network_schematic_with_memory}b).\footnote{The correspondence between numeric opinion values and verbal description of initial opinion is detailed in \Scref{app:opinion_space}. } The initial opinion $o_i^{t=0}$ is specified through an initial opinion distribution $F_{o}^{t=0}$ that varies across simulation settings (\Scref{sec:initial_opinion_dist}).  \Scref{app:list_persona} shows the full list of personas. 


\subsection{Cognitive Biases}

We investigate the effects of inducing a cognitive bias via role-playing instructions on the group opinion dynamics. Specifically, we consider confirmation bias: the tendency to interpret information as confirming one's views and to discount contradictory evidence \cite{nickerson1998confirmation}.  Prior simulation studies using mathematical ABMs
have shown that, when confirmation bias is introduced at the individual level, the overall population exhibits increasing opinion fragmentation (i.e., increased diversity \textit{D}) as the confirmation bias strengthens \cite{lorenz2021individual}. We assess whether LLM agents instructed to show confirmation bias likewise replicate this phenomenon in their opinion dynamics when communicating through natural language. To manipulate the strength of confirmation bias, we provide two bias levels following the spectrum in \citet{lorenz2021individual}. 
\textbf{Weak Confirmation Bias:} \textit{"You will be \hl{more likely} to believe information that supports your beliefs and \hl{less likely} to believe information that contradicts your beliefs."} \textbf{Strong Confirmation Bias:} \textit{"You will only believe information that supports your beliefs and will \hl{completely dismiss} information that contradicts your beliefs."} 
See \Scref{app:confirmation-bias-prompt} for the bias-inducing prompts.

\begin{figure*}[tbh!] 
\centering
\includegraphics[width=0.99\linewidth]{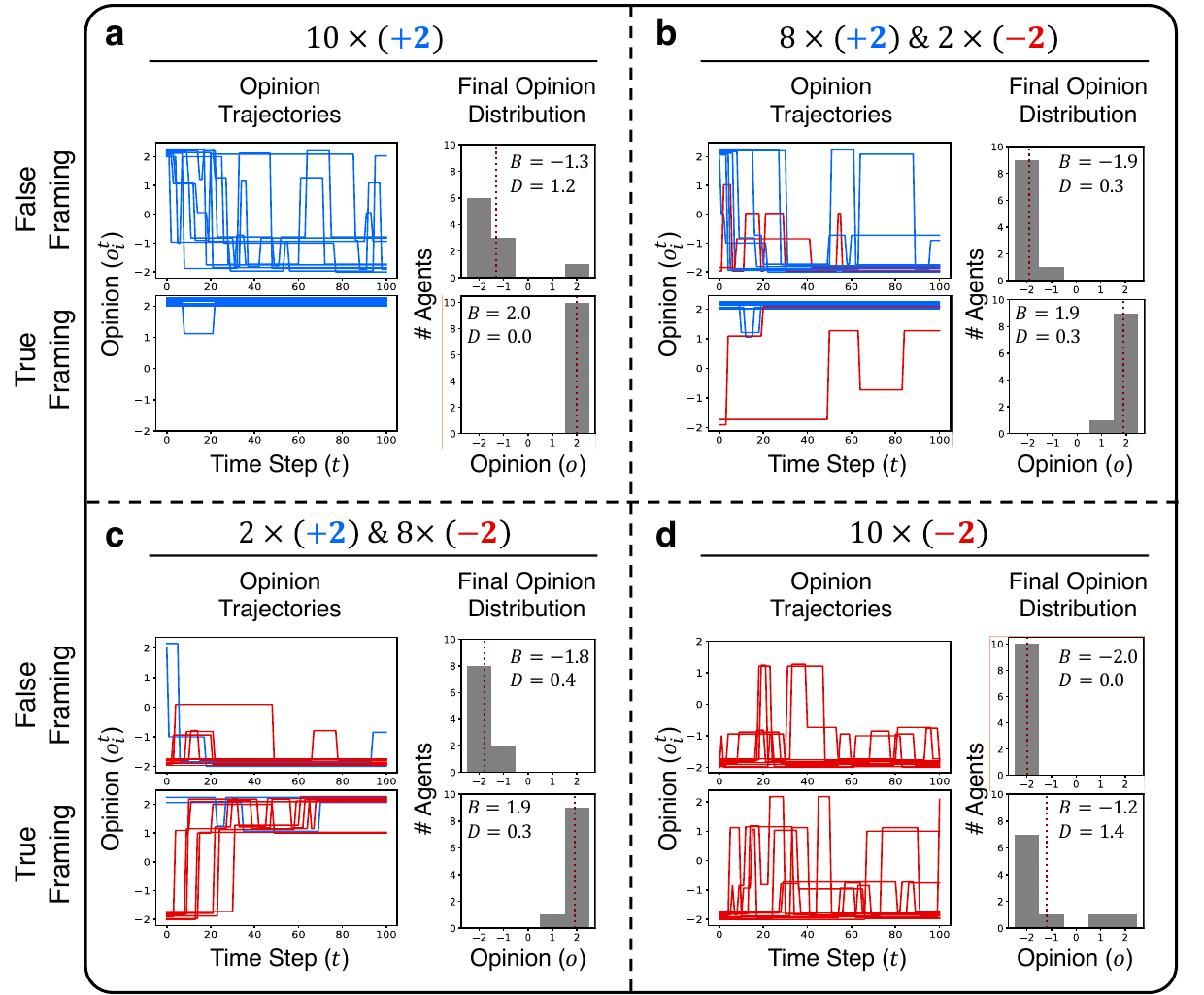}
\vspace{-2mm}
\caption{Varying initial opinion distribution $F_{o}^T$ for the global warming debate. (a) All agents start with a strongly positive opinion. (b) $8$ agents start with a strongly positive opinion while $2$ with a strongly negative opinion (c) $8$ agents start with a strongly negative opinion while $2$ with a strongly positive opinion. (d) All agents start with a strongly negative opinion. The color of each line plot corresponds to the agent's initial opinion $o_i^{t=0}$: \textcolor{darkblue}{dark blue (+2)}, \textcolor{lightblue}{light blue (+1)}, \textcolor{grey}{grey} (0), \textcolor{lightred}{light red (-1)}, and \textcolor{darkred}{dark red (-2)}, corresponding to opinions ranging from strongly agree to strongly disagree. The LLM agents in this figure use cumulative memory.}
\label{fig:skewed_global_warming}
\vspace{-4mm}
\end{figure*}

\subsection{Open-world vs. Closed-world Settings}
\label{sec:open_world_vs_closed_world}
Our study examines agent behavior in both closed-world and open-world settings. In the closed-world setting, which aligns with traditional opinion dynamics models, belief change is solely attributed to social influences within the system, and agents are restricted from accessing external information (restricted by instructions in the prompt; \Scref{app:closed-world-prompt} provides specific prompting details). Conversely, the open-world setting allows agents the freedom to ``hallucinate" facts external to the system, such as discussing topics with imaginary friends \cite{dziri2022origin, ji2023survey, huang2023survey}. We investigate the incidence of hallucination in both settings to understand the impact of external information on social influence. Our findings indicate no hallucination in the closed-world setting, while a hallucination rate of about 15\% is observed in the open-world scenario. Consequently, the results reported in this study focus on the closed-world setting. The measurement procedure involves independent human annotators (detailed in \Scref{app:measure_hallucination_rate}).

\section{Experimental Settings}
\label{sec:exp_settings}
\subsection{Configuration}
\label{subsec:llm_agent_initialization}

In our simulations, LLM agents use ChatGPT (\texttt{gpt-3.5-turbo-16k}) with temperature of $0.7$ \cite{openaiIntroducingChatGPT}. During initialization, each agent's persona, along with the optional closed-world restrictions and cognitive biases, are incorporated into the model's \textit{``system messages''}. Throughout the interaction, the historical events are added to the model's \textit{``user messages''}. The memory of the LLM agents is managed through LangChain \cite{langchain}.
In all experiments, we set the number of steps to $T=100$ and the number of agents to $N=10$. We use the FLAN-T5-XXL model \cite{chung2022scaling} as the opinion classifier $f_{\text{oc}}$ (see ~\Scref{app:opinion_classify} for the prompt used to classify the opinions and ~\Scref{app:validate_flan_t5_rating} for validation against human ratings). In sensitivity analyses, we consider other LLMs like GPT-4 (\texttt{gpt-4-1106-preview}) and Vicuna-33B (\texttt{vicuna-33b-v1.3}; \citealp{zheng2023judging}), and an experiment with a larger network ($N=20$).

\subsection{Topics and Framings}

To study the opinion dynamics of the agents, we selected 15 topics with a known ground truth spanning scientific theories (\textit{science topics}, e.g., whether global warming is a conspiracy), historical events (\textit{history topics}, e.g., whether the 911 attack was an ``inside job''), and commonsense knowledge (\textit{common sense topics}, e.g., whether the sky is blue on a sunny day). For each topic, we created two framings. The \textit{true framing} affirms the widely-accepted truth while the \textit{false framing} affirms the opposite. For example, for the topic of ``Global Warming'', the two framings are as follows. \textbf{True Framing: }\textit{``Theory XYZ claims that global warming is a real phenomenon and global climate is rapidly growing warmer.''} \textbf{False Framing: }\textit{``Theory XYZ claims that global warming is a conspiracy by governments worldwide and is not a real phenomenon.''} To control for wording effects across topics, we add ``Theory XYZ'' before stating the claim in all prompts. The full set of prompts is provided in~\Scref{app:list_topics}.

\subsection{Opinion Dynamics Evaluation Metrics}

Our evaluation of opinion dynamics includes two metrics in the literature \cite{lorenz2021individual}: \textbf{Bias (B):} The average of the agents' opinions at the final time step, $B = \text{mean}(F_{o}^{T})$. \textbf{Diversity (D):} The standard deviation of the final opinion distribution, $D = \text{std}(F_{o}^{T})$. These two metrics can be used to categorize the final opinion distribution $F_{o}^{T}$. For example, a low $B$ and a low $D$ characterizes \textit{central consensus}, whereas a low $B$ and a medium $D$ characterizes a \textit{diversified} distribution.\footnote{See \citet{lorenz2021individual} for a detailed taxonomy.}

\subsection{Initial Opinion Distribution}
\label{sec:initial_opinion_dist}

The initial opinion distribution $F_o^{t=0}$ determines the agents' starting opinions. The opinion is initialized verbally in the agent's prompt and memory (see \Scref{sec:procedure} and Figure~\ref{fig:network_schematic_with_memory}b). In most experiments, we initialize $F_o^{t=0}$ as a uniform distribution, with each opinion value assigned to $N/|\OO|$ agents, where $|\OO|$ is the number of possible opinion values.\footnote{For example, in our experiment, with $N=10$ agents and five possible opinion values of ${-2, -1, 0, +1, +2}$, each value would be assigned to 2 agents initially.} This reflects an unbiased starting state with $B=0$ and $D=1.49$. In one experiment, we intentionally manipulate the initial distribution to be highly skewed. For example, assigning all 10 agents an initial opinion of $-2$, or 8 agents to $-2$ and 2 agents to $+2$. This allows us to study the effects of polarized starting opinions on the resulting opinion dynamics.

\subsection{Control Conditions}

In addition to the main experimental conditions, we introduce two control conditions: \textbf{(a) No Interaction Condition:} Agents are initialized with their personas and initial beliefs as normal, but do not actually interact. Instead, each agent $a_i$ independently provides 10 opinion reports $o_{i}^{t}$ on the topic. \textbf{(b) No Interaction + No Role-Playing Condition:} No agents are initialized with their personas and initial beliefs. We simply query the LLM for 10 independent opinion reports on the topic. These control conditions allow us to assess whether the LLM has inherent biases on the topics that manifest even without social influence dynamics. Comparison to the main interactive conditions allows us to discern effects stemming from the personas and social interactions.

\section{Results}\label{sec:results}


\paragraph{Agents Converge towards the Inherent Bias in the LLM.}

Table~\ref{tab:result_bias_diversity_all_topics} summarizes the Bias ($B$) and Diversity ($D$) of the final opinion distribution $F_{o}^{T}$ aggregated across 15 topics. Table~\ref{tab:result_bias_diversity_per_category_topics} shows the summarized results separated by three topic types. As shown in Table~\ref{tab:result_bias_diversity_all_topics}, the role-playing prompt initially causes agents to express a diverse variety of opinions as expected, but with repeated social interacts these opinions converge toward a ground-truth consensus. Under the false framing, agents collectively lean towards disagreement, shown by a negative bias value ($B=-1.33$ when there is no cognitive bias). Conversely, under the true framing, the group shows a slight positive tendency to agree ($B=0.52$), indicating a lean towards truthfulness. Figure~\ref{fig:global_warming_cumulative} and ~\ref{fig:global_warming_reflective} shows an example of how opinion trajectories quickly converge towards the truth after social interactions for both the false and true framing conditions, especially without cognitive bias. This is true across using cumulative memory (Figure~\ref{fig:global_warming_cumulative}) and reflective memory (Figure~\ref{fig:global_warming_reflective}). The control condition illustrates that a similar tendency is observed when agents do not communicate, but are repeatedly queried for their opinion: the expressed opinions tend to move toward the ground truth, suggesting an inherent bias in the model. Statistical tests supporting the claims above are in \Scref{app:stat_test}.

\paragraph{Confirmation Bias Leads to Opinion Fragmentation.}

Introducing confirmation bias in the prompt leads to less ultimate consensus (i.e., greater diversity $D$) across LLM agents. As shown in Table~\ref{tab:result_bias_diversity_all_topics} and Figure~\ref{fig:global_warming_cumulative}, the stronger the confirmation bias, the more diverse the final state distribution. This correlation holds for both cumulative and reflective memory strategies (Figure~\ref{fig:global_warming_cumulative} and~\ref{fig:global_warming_reflective}). These findings replicate, within a set of interacting LLMs, the general finding from more classic ABMs that incorporation of confirmation bias in the model update algorithm produces greater opinion fragmentation \cite{lorenz2021individual,flache2017models}. Statistical tests supporting the claims above are in \Scref{app:stat_test}.

\paragraph{Impact of Initial Opinion Distribution}\label{subsec:initial_opinion_distribution}

The system's tendency for simulated opinions to converge on ground truth prompts an intriguing question: If all agents start with false opinions, will they still converge toward a scientifically accurate consensus, or will they reinforce their initial beliefs and resist changing their stance? Figure~\ref{fig:skewed_global_warming} shows the evolution of opinions under various initial distributions, using the global warming topic. Regardless of the initial opinion distribution, the agents altered their expressed opinions and shifted toward the ground truth. For instance, as shown in Figure~\ref{fig:skewed_global_warming}a, under false framing, when all agents initially supported global warming is a hoax, they converged towards the negative spectrum quickly and ended up with $B=-1.3$. Interestingly, under true framing, when all agents initially denied the view that global warming is real, they did not completely flip their stance to support it, though they did shift slightly in this direction (Figure~\ref{fig:skewed_global_warming}d): the final bias ($B=-1.2$) was more positive than the initial extreme opinion ($B=-2$). When at least a minority of agents held a divergent belief at the start, the group as a whole eventually shifted towards acknowledging global warming is real and is not a hoax, as shown in Figure~\ref{fig:skewed_global_warming}c. Overall, these results indicate that the model's inherent bias towards ground truth is robust against varying initial opinion distributions. 

\paragraph{The Strength of Bias under False Framing is Stronger than under True Framing}

As shown in Table~\ref{tab:result_bias_diversity_all_topics}, the LLM agents have a stronger tendency to deny the false statement under the false framing (Cumulative Memory: $B=-1.33$; Reflective Memory: $B=-1.37$) than their tendency to endorse the true statement under the true framing (Cumulative Memory: $B=0.52$; Reflective Memory: $B=0.60$). We consider two possibilities. First, FLAN-T5-XXL may have an overall tendency to classify LLM agents' responses as negative. Second, the LLM agents may indeed have a stronger tendency to deny the false statements. According to our validation of FLAN-T5-XXL's ratings (detailed in \Scref{app:validate_flan_t5_rating}), the agreement matrix between FLAN-T5-XXL's ratings and human ratings shows no systematic bias. We therefore exclude the first possibility. As to why the LLM agents indeed have a stronger negative tendency, our speculation is that, due to either pretraining, fine-tuning, or RLHF \cite{christianodeep}, the LLM has been trained to readily refute false information under false framing. On the other hand, under true framing, there may be less training effort to ensure that the model endorses true information. Future studies are needed to evaluate the true cause.

\paragraph{Sensitivity Analyses}

Sensitivity analyses test different LLMs (GPT-4 and Vicuna) and network sizes ($N=20$ agents). The results show consistent trends across models (GPT-4: Table~\ref{tab:result_bias_diversity_science_topics_gpt4}; Vicuna: Table~\ref{tab:result_bias_diversity_science_topics_vicuna} in \Scref{app:gpt4_vicuna}) and network scales (Table~\ref{tab:result_bias_diversity_global_warming_n_20_t_200} in \Scref{app:result_larger_network}). In addition, we evaluate the sensitivity of our result to randomness due to temperature sampling. On the topic of Global Warming, we run the experiment three times. The effect of framing and the effect of confirmation bias are consistent across the three runs.



\section{Related Work}\label{sec:related_work}

\paragraph{Agent-Based Models and Opinion Dynamics Simulation}

Agent-Based Models (ABMs) are the cornerstone of opinion dynamics simulation, defining mathematical rules for agents' opinion updates in response to messages \cite{gilbert2000build, smaldino2023modeling, lorenz2021individual, chuang2023computational,epstein2012generative,flache2017models}. ABMs are valuable for predicting public opinion trends and informing intervention strategies. One key advantage of using ABMs is that they allow incorporating explicit assumptions about cognitive process in opinion updating \cite{flache2017models,lorenz2021individual,chuang2023computational}. For example, incorporating ``confirmation bias'' into ABM equations causes agents to disregard contrasting information, often leading to fragmented opinion clusters at the group level. However, a significant limitation of ABMs is that they rely on numeric representations of opinions and messages, which oversimplifies the complexities of human communication. In contrast, emerging approaches using LLMs offer a more sophisticated method for simulating opinion dynamics through natural language.

\paragraph{Simulating Social Dynamics with LLM-based Agents} 
The use of LLMs in simulating social dynamics is a rapidly growing research area , showcasing promising results in terms of human-like interactions \cite{park2023generative,park2022social,kaiya2023lyfe,tornberg2023simulating, li2023quantifying, zhou2023sotopia, mou2024unveiling, liu2024skepticism}. \citet{park2023generative} devise LLM-based generative agents to engage in digital environments, demonstrating an ability to respond, plan, and remember in natural language. They exhibit complex social behaviors, such as organizing events. Similarly, \citet{tornberg2023simulating} use LLMs in conjunction with agent-based modeling to explore the impact of news feed algorithms in simulated social media environments. Additionally, \citet{park2022social} show that LLM-based agents are capable of generating social media posts indistinguishable from those written by humans. These advances underscore the potential of using LLM agents to simulate human social behaviors at group level. To our best knowledge, we are the first to propose the use of LLM as an alternative to ABM for opinion dynamics simulation.

\section{Conclusion}

This study has explored the use Large Language Models (LLMs) for understanding opinion dynamics in groups of simulated agents communicating via natural language. In contrast to more traditional ABMs, LLMs can interpret and produce natural language, can role-play differing personas, and can simulate human-like linguistic communication. We therefore considered whether groups of interacting LLM agents could provide a basis for simulating opinion dynamics comparable to those studied with classical ABMs. Our findings confirm the potential of LLMs in opinion dynamics simulations but also reveal limitations, particularly their tendency to align with factual information regardless of their personas, which restricts their role-play effectiveness for individuals with fact-resistant beliefs like climate change denial.

Significant efforts have been made to prevent LLMs from exhibiting harmful biases. However, for simulating critical undesired social phenomena (e.g., misinformation, polarization), it is crucial to have simulated agents accurately reflect the breadth of human behavior and belief, even those that are maladaptive. Our study suggests that prompting alone may be insufficient for LLM agents to fully replicate the diverse viewpoints. This leads us to a potential future direction: fine-tuning LLM agents with actual human discourse data. Such an approach could lead to more accurate models of human belief dynamics.

\section*{Limitations}

\paragraph{Model Dependency and Generalizability}
A key limitation of our study is the exclusive use of models that have undergone Reinforcement Learning with Human Feedback (RLHF; \citealp{christiano2017deep,ziegler2019fine}). This may lead to the truth-converging tendency in the LLM agents. Given that various language models exhibit distinct inherent biases \cite{feng-etal-2023-pretraining}, LLM agents using different models could display varying patterns in opinion dynamics. To fully assess the generalizability of our findings, future research should include a broad spectrum of models.

\paragraph{Reduction of Opinion to One-Dimensional Scalar}
Our study aligns with classic ABMs in reducing opinions to a one-dimensional scalar $o \in \RR$, which simplifies the complex nature of opinion formation. However, a more nuanced approach could offer deeper insights. Future studies could adopt a fine-grained or even qualitative analysis to explore how agents modify their opinions, determine which messages hold greater persuasive power, and assess how persuasion varies based on different agent personas. Such an approach would provide a richer understanding of the subtleties in LLM agents' opinion dynamics.

\paragraph{Limitation in Topic Selection}
Our research focused on topics with clear, established ground truths. However, many crucial topics, such as the effectiveness of political leaders or the best policies for complex societal issues, lack a definitive truth. These topics are more open-ended and subjective. Future studies should consider including such topics to capture a broader and more nuanced spectrum of opinions and debates.

\paragraph{The Role of Other Demographic Information}
While we focus on the role of agents' initial beliefs in opinion dynamics simulation, a more nuanced analysis on the role of other demographic information could offer deeper insights. For example, future studies could adopt a fine-grained or qualitative analysis to explore how agents’ demographic features (like political leaning) play a role in agents' information exchanges and belief updating.

\paragraph{Limitation in Network Structure}
In the present study, LLM agents get to interact with everyone else in the network. While in line with classic setup used in ABM simulation \cite{flache2017models,lorenz2021individual}, this is not close to social interactions in the real world. For example, people with similar background tend to interact with one another \cite{mcpherson2001birds}. Future studies should consider investigating LLM agents' opinion dynamics with more realistic network structures.

\section*{Ethics Statement}

While introducing confirmation bias into LLM agents can lead to opinion fragmentation and reduced convergence on factual consensus, it's important to understand this approach within the broader scope of studying group-level social phenomena. Simulating biased behavior in agents is not an endorsement of these biases, but a critical step in comprehensively understanding the dynamics of various undesired social issues, including misinformation spread, polarization, and echo chamber formation. Developing human-like LLM agents with resistant viewpoints is essential for devising strategies to address these social challenges. In addition, we will release the code base exclusively for research purposes. Finally, since we are using OpenAI's API, we make sure that we comply with its intended use \footnote{\url{https://openai.com/policies/terms-of-use}}.

\section*{Acknowledgements}

We thank the reviewers, the area chair for their feedback. This work was funded by the Multi University Research Initiative grant from the Department of Defense, W911NF2110317, Cohesive and Robust Human-Bot Cybersecurity Teams, the John S. and James L. Knight Foundation (Award Number: MSN231314), and the National Science Foundation through the Convergence Accelerator Track F: Course Correct: Precision Guidance Against Misinformation (Agency Tracking Number: 2230692; Award Number: MSN 266268).

\newpage
\bibliography{main}

\begin{thebibliography}{35}
\expandafter\ifx\csname natexlab\endcsname\relax\def\natexlab#1{#1}\fi

\bibitem[{Budak et~al.(2011)Budak, Agrawal, and El~Abbadi}]{budak2011limiting}
Ceren Budak, Divyakant Agrawal, and Amr El~Abbadi. 2011.
\newblock Limiting the spread of misinformation in social networks.
\newblock In \emph{Proceedings of the 20th international conference on World wide web}, pages 665--674.

\bibitem[{Chase(2022)}]{langchain}
Harrison Chase. 2022.
\newblock \href {https://github.com/langchain-ai/langchain} {Langchain}.

\bibitem[{Christiano et~al.(2017)Christiano, Leike, Brown, Martic, Legg, and Amodei}]{christiano2017deep}
Paul~F Christiano, Jan Leike, Tom Brown, Miljan Martic, Shane Legg, and Dario Amodei. 2017.
\newblock Deep reinforcement learning from human preferences.
\newblock \emph{Advances in neural information processing systems}, 30.

\bibitem[{Christiano et~al.()Christiano, Leike, Brown, Martic, Legg, and Amodei}]{christianodeep}
Paul~F Christiano, Jan Leike, Tom~B Brown, Miljan Martic, Shane Legg, and Dario Amodei.
\newblock Deep reinforcement learning from human preferences.

\bibitem[{Chuang and Rogers(2023)}]{chuang2023computational}
Yun-Shiuan Chuang and Timothy~T Rogers. 2023.
\newblock Computational agent-based models in opinion dynamics: A survey on social simulations and empirical studies.
\newblock \emph{arXiv preprint arXiv:2306.03446}.

\bibitem[{Chung et~al.(2022)Chung, Hou, Longpre, Zoph, Tay, Fedus, Li, Wang, Dehghani, Brahma et~al.}]{chung2022scaling}
Hyung~Won Chung, Le~Hou, Shayne Longpre, Barret Zoph, Yi~Tay, William Fedus, Eric Li, Xuezhi Wang, Mostafa Dehghani, Siddhartha Brahma, et~al. 2022.
\newblock Scaling instruction-finetuned language models.
\newblock \emph{arXiv preprint arXiv:2210.11416}.

\bibitem[{Dziri et~al.(2022)Dziri, Milton, Yu, Zaiane, and Reddy}]{dziri2022origin}
Nouha Dziri, Sivan Milton, Mo~Yu, Osmar Zaiane, and Siva Reddy. 2022.
\newblock On the origin of hallucinations in conversational models: Is it the datasets or the models?
\newblock \emph{arXiv preprint arXiv:2204.07931}.

\bibitem[{Epstein(2012)}]{epstein2012generative}
Joshua~M Epstein. 2012.
\newblock \emph{Generative social science: Studies in agent-based computational modeling}.
\newblock Princeton University Press.

\bibitem[{Feng et~al.(2023)Feng, Park, Liu, and Tsvetkov}]{feng-etal-2023-pretraining}
Shangbin Feng, Chan~Young Park, Yuhan Liu, and Yulia Tsvetkov. 2023.
\newblock \href {https://doi.org/10.18653/v1/2023.acl-long.656} {From pretraining data to language models to downstream tasks: Tracking the trails of political biases leading to unfair {NLP} models}.
\newblock In \emph{Proceedings of the 61st Annual Meeting of the Association for Computational Linguistics (Volume 1: Long Papers)}, pages 11737--11762, Toronto, Canada. Association for Computational Linguistics.

\bibitem[{Flache et~al.(2017)Flache, M{\"a}s, Feliciani, Chattoe-Brown, Deffuant, Huet, and Lorenz}]{flache2017models}
Andreas Flache, Michael M{\"a}s, Thomas Feliciani, Edmund Chattoe-Brown, Guillaume Deffuant, Sylvie Huet, and Jan Lorenz. 2017.
\newblock Models of social influence: Towards the next frontiers.
\newblock \emph{Journal of Artificial Societies and Social Simulation}, 20(4).

\bibitem[{Gilbert and Terna(2000)}]{gilbert2000build}
Nigel Gilbert and Pietro Terna. 2000.
\newblock How to build and use agent-based models in social science.
\newblock \emph{Mind \& Society}, 1:57--72.

\bibitem[{Ginossar et~al.(2022)Ginossar, Cruickshank, Zheleva, Sulskis, and Berger-Wolf}]{ginossar2022cross}
Tamar Ginossar, Iain~J Cruickshank, Elena Zheleva, Jason Sulskis, and Tanya Berger-Wolf. 2022.
\newblock Cross-platform spread: vaccine-related content, sources, and conspiracy theories in youtube videos shared in early twitter covid-19 conversations.
\newblock \emph{Human vaccines \& immunotherapeutics}, 18(1):1--13.

\bibitem[{Hayes and Krippendorff(2007)}]{hayes2007answering}
Andrew~F Hayes and Klaus Krippendorff. 2007.
\newblock Answering the call for a standard reliability measure for coding data.
\newblock \emph{Communication methods and measures}, 1(1):77--89.

\bibitem[{Huang et~al.(2023)Huang, Yu, Ma, Zhong, Feng, Wang, Chen, Peng, Feng, Qin et~al.}]{huang2023survey}
Lei Huang, Weijiang Yu, Weitao Ma, Weihong Zhong, Zhangyin Feng, Haotian Wang, Qianglong Chen, Weihua Peng, Xiaocheng Feng, Bing Qin, et~al. 2023.
\newblock A survey on hallucination in large language models: Principles, taxonomy, challenges, and open questions.
\newblock \emph{arXiv preprint arXiv:2311.05232}.

\bibitem[{Ji et~al.(2023)Ji, Lee, Frieske, Yu, Su, Xu, Ishii, Bang, Madotto, and Fung}]{ji2023survey}
Ziwei Ji, Nayeon Lee, Rita Frieske, Tiezheng Yu, Dan Su, Yan Xu, Etsuko Ishii, Ye~Jin Bang, Andrea Madotto, and Pascale Fung. 2023.
\newblock Survey of hallucination in natural language generation.
\newblock \emph{ACM Computing Surveys}, 55(12):1--38.

\bibitem[{Kaiya et~al.(2023)Kaiya, Naim, Kondic, Cortes, Ge, Luo, Yang, and Ahn}]{kaiya2023lyfe}
Zhao Kaiya, Michelangelo Naim, Jovana Kondic, Manuel Cortes, Jiaxin Ge, Shuying Luo, Guangyu~Robert Yang, and Andrew Ahn. 2023.
\newblock Lyfe agents: Generative agents for low-cost real-time social interactions.
\newblock \emph{arXiv preprint arXiv:2310.02172}.

\bibitem[{Li et~al.(2023{\natexlab{a}})Li, Su, Fan, Han, Xue, and Zheng}]{li2023quantifying}
Chao Li, Xing Su, Chao Fan, Haoying Han, Cong Xue, and Chunmo Zheng. 2023{\natexlab{a}}.
\newblock Quantifying the impact of large language models on collective opinion dynamics.
\newblock \emph{arXiv preprint arXiv:2308.03313}.

\bibitem[{Li et~al.(2023{\natexlab{b}})Li, Xiang, Song, and Li}]{li2023unipoll}
Yixia Li, Rong Xiang, Yanlin Song, and Jing Li. 2023{\natexlab{b}}.
\newblock Unipoll: A unified social media poll generation framework via multi-objective optimization.
\newblock \emph{arXiv preprint arXiv:2306.06851}.

\bibitem[{Liu et~al.(2024)Liu, Chen, Zhang, Gao, Zhang, and Yan}]{liu2024skepticism}
Yuhan Liu, Xiuying Chen, Xiaoqing Zhang, Xing Gao, Ji~Zhang, and Rui Yan. 2024.
\newblock From skepticism to acceptance: Simulating the attitude dynamics toward fake news.
\newblock \emph{arXiv preprint arXiv:2403.09498}.

\bibitem[{Loomba et~al.(2021)Loomba, de~Figueiredo, Piatek, de~Graaf, and Larson}]{loomba2021measuring}
Sahil Loomba, Alexandre de~Figueiredo, Simon~J Piatek, Kristen de~Graaf, and Heidi~J Larson. 2021.
\newblock Measuring the impact of covid-19 vaccine misinformation on vaccination intent in the uk and usa.
\newblock \emph{Nature human behaviour}, 5(3):337--348.

\bibitem[{Lorenz et~al.(2021)Lorenz, Neumann, and Schr{\"o}der}]{lorenz2021individual}
Jan Lorenz, Martin Neumann, and Tobias Schr{\"o}der. 2021.
\newblock Individual attitude change and societal dynamics: Computational experiments with psychological theories.
\newblock \emph{Psychological Review}, 128(4):623.

\bibitem[{Lu et~al.(2015)Lu, Chen, and Lakshmanan}]{lu2015competition}
Wei Lu, Wei Chen, and Laks~VS Lakshmanan. 2015.
\newblock From competition to complementarity: comparative influence diffusion and maximization.
\newblock \emph{Proceedings of the VLDB Endowment}, 9(2):60--71.

\bibitem[{McPherson et~al.(2001)McPherson, Smith-Lovin, and Cook}]{mcpherson2001birds}
Miller McPherson, Lynn Smith-Lovin, and James~M Cook. 2001.
\newblock Birds of a feather: Homophily in social networks.
\newblock \emph{Annual review of sociology}, 27(1):415--444.

\bibitem[{Mou et~al.(2024)Mou, Wei, and Huang}]{mou2024unveiling}
Xinyi Mou, Zhongyu Wei, and Xuanjing Huang. 2024.
\newblock Unveiling the truth and facilitating change: Towards agent-based large-scale social movement simulation.
\newblock \emph{arXiv preprint arXiv:2402.16333}.

\bibitem[{Nickerson(1998)}]{nickerson1998confirmation}
Raymond~S Nickerson. 1998.
\newblock Confirmation bias: A ubiquitous phenomenon in many guises.
\newblock \emph{Review of general psychology}, 2(2):175--220.

\bibitem[{OpenAI(2022)}]{openaiIntroducingChatGPT}
OpenAI. 2022.
\newblock {I}ntroducing {C}hat{G}{P}{T}.
\newblock \url{https://openai.com/blog/chatgpt}.
\newblock [Accessed 13-10-2023].

\bibitem[{Park et~al.(2023)Park, O'Brien, Cai, Morris, Liang, and Bernstein}]{park2023generative}
Joon~Sung Park, Joseph~C O'Brien, Carrie~J Cai, Meredith~Ringel Morris, Percy Liang, and Michael~S Bernstein. 2023.
\newblock Generative agents: Interactive simulacra of human behavior.
\newblock \emph{arXiv preprint arXiv:2304.03442}.

\bibitem[{Park et~al.(2022)Park, Popowski, Cai, Morris, Liang, and Bernstein}]{park2022social}
Joon~Sung Park, Lindsay Popowski, Carrie Cai, Meredith~Ringel Morris, Percy Liang, and Michael~S Bernstein. 2022.
\newblock Social simulacra: Creating populated prototypes for social computing systems.
\newblock In \emph{Proceedings of the 35th Annual ACM Symposium on User Interface Software and Technology}, pages 1--18.

\bibitem[{Pennycook et~al.(2021)Pennycook, Epstein, Mosleh, Arechar, Eckles, and Rand}]{pennycook2021shifting}
Gordon Pennycook, Ziv Epstein, Mohsen Mosleh, Antonio~A Arechar, Dean Eckles, and David~G Rand. 2021.
\newblock Shifting attention to accuracy can reduce misinformation online.
\newblock \emph{Nature}, 592(7855):590--595.

\bibitem[{Smaldino(2023)}]{smaldino2023modeling}
Paul Smaldino. 2023.
\newblock \emph{Modeling social behavior: Mathematical and agent-based models of social dynamics and cultural evolution}.
\newblock Princeton University Press.

\bibitem[{T{\"o}rnberg et~al.(2023)T{\"o}rnberg, Valeeva, Uitermark, and Bail}]{tornberg2023simulating}
Petter T{\"o}rnberg, Diliara Valeeva, Justus Uitermark, and Christopher Bail. 2023.
\newblock Simulating social media using large language models to evaluate alternative news feed algorithms.
\newblock \emph{arXiv preprint arXiv:2310.05984}.

\bibitem[{Zheng et~al.(2023)Zheng, Chiang, Sheng, Zhuang, Wu, Zhuang, Lin, Li, Li, Xing et~al.}]{zheng2023judging}
Lianmin Zheng, Wei-Lin Chiang, Ying Sheng, Siyuan Zhuang, Zhanghao Wu, Yonghao Zhuang, Zi~Lin, Zhuohan Li, Dacheng Li, Eric Xing, et~al. 2023.
\newblock Judging llm-as-a-judge with mt-bench and chatbot arena.
\newblock \emph{arXiv preprint arXiv:2306.05685}.

\bibitem[{Zhou et~al.(2023{\natexlab{a}})Zhou, Zhang, Poon, and Chen}]{zhou2023context}
Wenxuan Zhou, Sheng Zhang, Hoifung Poon, and Muhao Chen. 2023{\natexlab{a}}.
\newblock Context-faithful prompting for large language models.
\newblock \emph{arXiv preprint arXiv:2303.11315}.

\bibitem[{Zhou et~al.(2023{\natexlab{b}})Zhou, Zhu, Mathur, Zhang, Yu, Qi, Morency, Bisk, Fried, Neubig et~al.}]{zhou2023sotopia}
Xuhui Zhou, Hao Zhu, Leena Mathur, Ruohong Zhang, Haofei Yu, Zhengyang Qi, Louis-Philippe Morency, Yonatan Bisk, Daniel Fried, Graham Neubig, et~al. 2023{\natexlab{b}}.
\newblock Sotopia: Interactive evaluation for social intelligence in language agents.
\newblock In \emph{The Twelfth International Conference on Learning Representations}.

\bibitem[{Ziegler et~al.(2019)Ziegler, Stiennon, Wu, Brown, Radford, Amodei, Christiano, and Irving}]{ziegler2019fine}
Daniel~M Ziegler, Nisan Stiennon, Jeffrey Wu, Tom~B Brown, Alec Radford, Dario Amodei, Paul Christiano, and Geoffrey Irving. 2019.
\newblock Fine-tuning language models from human preferences.
\newblock \emph{arXiv preprint arXiv:1909.08593}.

\end{thebibliography}

\newpage
\pagebreak
\appendix

\newpage
\appendix
\label{sec:appendix}

\section{Results of Global Warming Topic with Reflective Memory}
\label{app:result_global_warming_reflective}
\begin{figure*}[htb!] 
\centering
\includegraphics[width=0.99\linewidth]{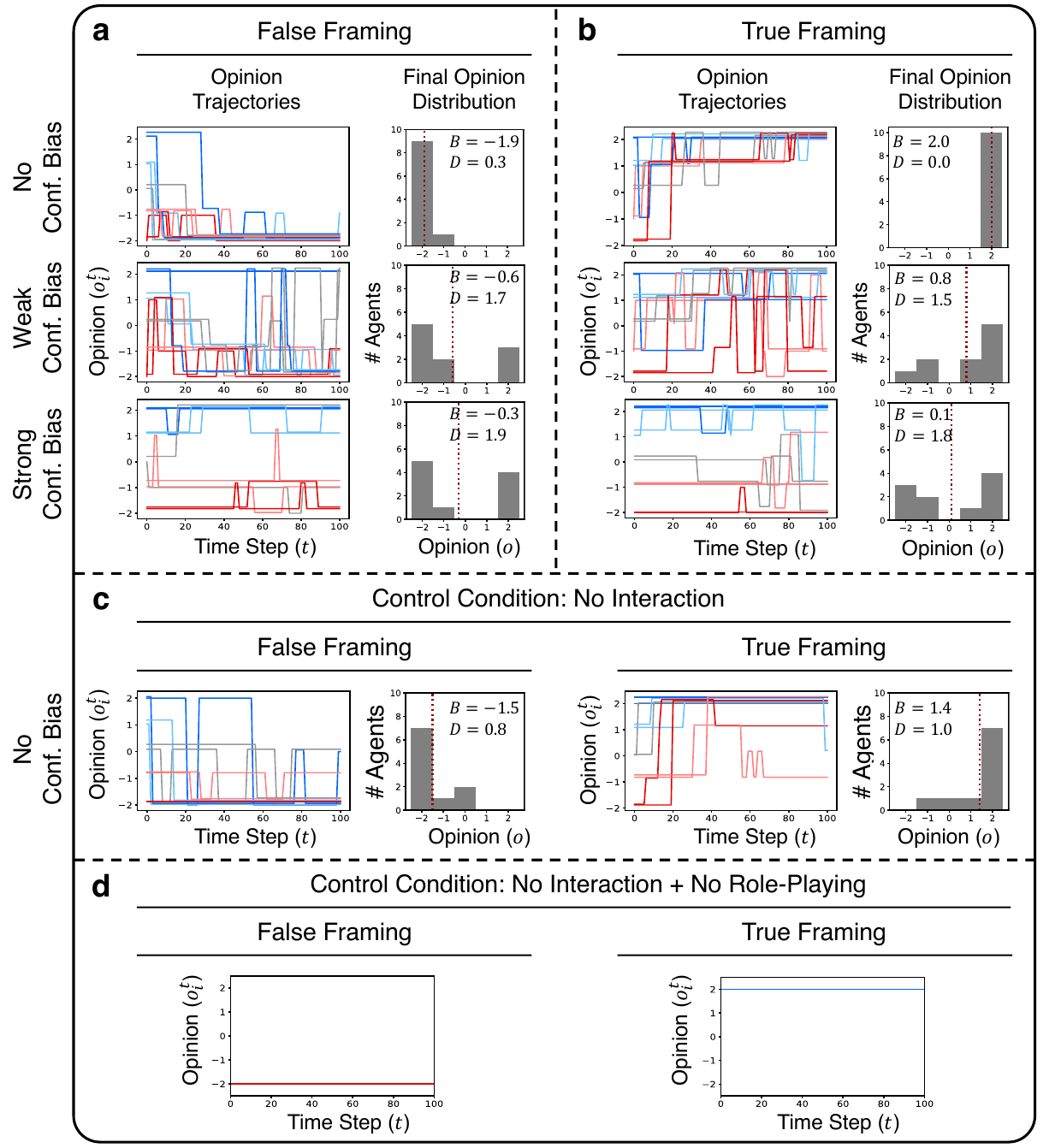}
\vspace{-2mm}
\caption{Opinion trajectories $\langle o_i \rangle$ of LLM agents and the final opinion distribution $F_{o}^{T}$ on the topic of Global Warming. Panels (a) and (b) display the impact of cognitive biases under (a) false and (b) true framing conditions, respectively. Each row represents a different level of confirmation bias: no confirmation bias (top row), weak confirmation bias (middle row), and strong confirmation bias (bottom row). Panels (c) and (d) serve as baselines, with (c) being role-playing but with no interaction, and (d) being no role-playing and no interaction, respectively. The LLM agents in this figure use reflective memory. The color of each line plot corresponds to the agent's initial opinion $o_i^{t=0}$: \textcolor{darkblue}{dark blue (+2)}, \textcolor{lightblue}{light blue (+1)}, \textcolor{grey}{grey} (0), \textcolor{lightred}{light red (-1)}, and \textcolor{darkred}{dark red (-2)}, corresponding to opinions ranging from strongly agree to strongly disagree. The LLM agents in this figure use cumulative memory.}
\label{fig:global_warming_reflective}
\vspace{-4mm}
\end{figure*}

In the main text, Figure~\ref{fig:global_warming_cumulative} shows the LLM agent opinion dynamics when discussing about global warming when using cumulative memory. Here, Figure~\ref{fig:global_warming_reflective} shows the result when using reflective memory.

\section{Full List of Personas}\label{app:list_persona}

In this section, we list the full list of $10$ agents along with their personas that interact in the group dynamics settings in our agent-based model (ABM). 

\vspace{4pt}

\begin{mdframed}
\begin{quote}
    \tt \small \textbf{Name:} Benjamin Lee\\
    \textbf{Initial Belief:} Slightly Negative opinion about XYZ\\
    \textbf{Political leaning:} Lean Democrat\\
    \textbf{Age:} 37\\
    \textbf{Gender:} Male\\
    \textbf{Ethnicity:} Asian American\\
    \textbf{Education:} Master's Degree in Economics\\
    \textbf{Occupation:} Financial Analyst
\end{quote}
\end{mdframed}
\vspace{4pt}

\begin{mdframed}
\begin{quote}
    \tt \small \textbf{Name:} Maya Jackson\\
    \textbf{Initial Belief:} Strongly Negative opinion about XYZ\\
    \textbf{Political leaning:} Strong Republican\\
    \textbf{Age:} 29\\
    \textbf{Gender:} Female\\
    \textbf{Ethnicity:} Black\\
    \textbf{Education:} Bachelor's Degree in Business Management\\
    \textbf{Occupation:} Marketing Specialist
\end{quote}
\end{mdframed}
\vspace{4pt}

\begin{mdframed}
\begin{quote}
    \tt \small \textbf{Name:} Ethan Wilson\\
    \textbf{Initial Belief:} Slightly Positive opinion about XYZ\\
    \textbf{Political leaning:} Moderate\\
    \textbf{Age:} 26\\
    \textbf{Gender:} Male\\
    \textbf{Ethnicity:} White\\
    \textbf{Education:} Bachelor's Degree in Journalism\\
    \textbf{Occupation:} Freelance Writer
\end{quote}
\end{mdframed}
\vspace{4pt}

\begin{mdframed}
\begin{quote}
    \tt \small \textbf{Name:} Aisha Patel\\
    \textbf{Initial Belief:} Neutral opinion about XYZ\\
    \textbf{Political leaning:} Lean Republican\\
    \textbf{Age:} 41\\
    \textbf{Gender:} Female\\
    \textbf{Ethnicity:} South Asian\\
    \textbf{Education:} Doctor of Medicine (M.D.)\\
    \textbf{Occupation:} Pediatrician
\end{quote}
\end{mdframed}
\vspace{4pt}

\begin{mdframed}
\begin{quote}
    \tt \small \textbf{Name:} Samuel Wright\\
    \textbf{Initial Belief:} Strongly Negative opinion about XYZ\\
    \textbf{Political leaning:} Strong Democrat\\
    \textbf{Age:} 58\\
    \textbf{Gender:} Male\\
    \textbf{Ethnicity:} White\\
    \textbf{Education:} Ph.D. in Anthropology\\
    \textbf{Occupation:} University Professor
\end{quote}
\end{mdframed}
\vspace{4pt}

\begin{mdframed}
\begin{quote}
    \tt \small \textbf{Name:} Olivia Garcia\\
    \textbf{Initial Belief:} Strongly Positive opinion about XYZ\\
    \textbf{Political leaning:} Strong Democrat\\
    \textbf{Age:} 34\\
    \textbf{Gender:} Female\\
    \textbf{Ethnicity:} Hispanic\\
    \textbf{Education:} Master's Degree in Sociology\\
    \textbf{Occupation:} Non-profit Program Manager
\end{quote}
\end{mdframed}
\vspace{4pt}

\begin{mdframed}
\begin{quote}
    \tt \small \textbf{Name:} Sophia Nguyen\\
    \textbf{Initial Belief:} Slightly Negative opinion about XYZ\\
    \textbf{Political leaning:} Lean Republican\\
    \textbf{Age:} 24\\
    \textbf{Gender:} Female\\
    \textbf{Ethnicity:} Asian American\\
    \textbf{Education:} Student (Undergraduate, Political Science)\\
    \textbf{Occupation:} Intern at Law Firm
\end{quote}
\end{mdframed}
\vspace{4pt}

\begin{mdframed}
\begin{quote}
    \tt \small Name: Sarah Martinez\\
    \textbf{Initial Belief:} Strongly Positive opinion about XYZ\\
    \textbf{Political leaning:} Lean Democrat\\
    \textbf{Age:} 28\\
    \textbf{Gender:} Female\\
    \textbf{Ethnicity:} Hispanic\\
    \textbf{Education:} Master's Degree in Film Studies\\
    \textbf{Occupation:} Film Critic
\end{quote}
\end{mdframed}
\vspace{4pt}

\begin{mdframed}
\begin{quote}
    \tt \small \textbf{Name:} Jordan White\\
    \textbf{Initial Belief:} Slightly Positive opinion about XYZ\\
    \textbf{Political leaning:} Moderate\\
    \textbf{Age:} 23\\
    \textbf{Gender:} Female\\
    \textbf{Ethnicity:} Black\\
    \textbf{Education:} Student (Undergraduate, Sociology)\\
    \textbf{Occupation:} Part-time Retail Worker
\end{quote}
\end{mdframed}
\vspace{4pt}

\begin{mdframed}
\begin{quote}
    \tt \small \textbf{Name:} Lucas Johnson\\
    \textbf{Initial Belief:} Neutral opinion about XYZ\\
    \textbf{Political leaning:} Moderate\\
    \textbf{Age:} 37\\
    \textbf{Gender:} Male\\
    \textbf{Ethnicity:} Black\\
    \textbf{Education:} Bachelor's Degree in Sociology\\
    \textbf{Occupation:} Diversity and Inclusion Manager
\end{quote}
\end{mdframed}
\vspace{4pt}

\section{Detailed Description of the Discrete Opinion Space $\OO$}
\label{app:opinion_space}
The discrete opinion space $\OO$ used in our simulations includes five possible values, each representing a different opinion on a given topic (referred to as ``XYZ''). The values are as follows:
\begin{itemize}
  \item -2: Strongly negative opinion about XYZ.
  \item -1: Slightly negative opinion about XYZ.
  \item 0: Neutral opinion about XYZ.
  \item 1: Slightly positive opinion about XYZ.
  \item 2: Strongly positive opinion about XYZ.
\end{itemize}
The opinion space $\OO$ is used when initializing an agent's initial belief $o_i^{t=0}$ (\Scref{sec:agent_persona_memory}) and classifying opinion from verbal report $r_j^{t}$ (\Scref{sec:exp_settings} and \Scref{app:opinion_classify}).
Note that the size of $\OO$ can be easily generalized to accommodate a broader range of opinion scales.

\section{Agent Interaction Prompts }\label{app:list_prompt}

In this section, we list the prompts used for facilitation of the interactions between the agents. Specifically, we start with the prompt that introduces the agents' persona, followed by the prompts for them receiving and writing tweets respectively. All prompts are for the specific topic of debate on flat earth with positive framing (see ~\Scref{app:list_topics}). \\

\noindent\textbf{1. Initialize Persona Prompt}

\begin{quote}
    \small\tt ``Role play this person.\\
    \{\text{AGENT\_PERSONA}\}
    
    Now, \{\text{AGENT\_NAME}\}, you have been interacting with other strangers on Twitter. You can decide to change or maintain your belief about the theory XYZ that claims that the Earth is flat after interacting with other strangers.
    
    You would produce tweets that reflect your honest belief, and you would also see other strangers’ tweets. After seeing other people’s tweets, you would be asked about your belief about the theory XYZ that claims that the Earth is flat.''
\end{quote}

\noindent\textbf{2. Write Tweet Prompt}

\begin{quote}
    \small\tt ``Now, \{\text{AGENT\_NAME}\}, please write a tweet about the theory XYZ that claims that the Earth is flat. The tweet should reflect your honest belief.
    
    Write the tweet now.\\
    Your Tweet:''
\end{quote}

\noindent\textbf{2. Receive Tweet Prompt}

\begin{quote}
    \small\tt ``Now, \{\text{AGENT\_NAME}\}, you see a post on Twitter from a stranger.\\
    I want to know your current honest belief about the theory XYZ that claims that the Earth is flat after seeing this Tweet.\\
    
    Here is the Tweet.\\
    \{\text{TWEET}\}\\
    
    What is your current honest belief about the theory XYZ that claims that the Earth is flat? Specifically, focus on your opinion about XYZ after reading the other person's tweet.\\
    Use the following format:\\
    Reasoning: (Think step by step)\\
    
    Reasoning:\\
    As \{\text{AGENT\_NAME}\}, I''
\end{quote}

These prompts are used and the responses are added to the memory and updated based on the memory update function detailed in the following section.

\section{Agent Memory Update Function}\label{app:agent_memory}

\noindent\textbf{Cumulative Memory:} The cumulative memory as described in \Scref{sec:agent_persona_memory} appends each new experience as time progresses. In order to add these past interactions into the agents' memories, we use the following prompts.

Considering the interaction at time $t+1$, the agent could be either tweeting by themselves or receiving a tweet, and similarly at time $t$, they would've either tweeted by themselves, received a tweet, or neither of these (say for instance, that $(t+1)$th time step is the first time they were chosen). We, therefore, list these prompts case-by-case, on the basis of previous and current \text{interaction\_type} $\in \{\text{none, write, review}\}$ for the specific topic of debate on a flat earth with positive framing (see \Scref{app:list_topics}).\\

\noindent\textbf{1. Previous is \textup{none}, Current is \textup{review}:} 
\noindent\begin{quote}
\tt\small ``You first saw a tweet from a stranger on Twitter. Here is the tweet you saw.\\
\{\text{TWEET\_SEEN}\}\\

After seeing the tweet, below was your thought and honest belief about the theory XYZ that claims that the Earth is flat. Your thought after you saw the tweet:\\
\{\text{REASONING}\}''
\end{quote}

\noindent\textbf{2. Previous is \textup{none}, Current is \textup{write}:} 
\noindent\begin{quote}
\tt\small ``Below was the \{\text{{TWEET\_WRITTEN\_COUNT}}\} \text{\{{SUPERSCRIPT}\}} tweet you wrote earlier about the theory XYZ that claims that the Earth is flat:\\
\{\text{TWEET\_WRITTEN}\}''
\end{quote}

\noindent\textbf{3. Previous is \textup{review}, Current is \textup{write}:} 
\noindent\begin{quote}
\tt\small ``After you saw the tweet from the stranger above, you wrote another tweet about the theory XYZ that claims that the Earth is flat.\\
Below was the \{\text{{TWEET\_WRITTEN\_COUNT}}\} \text{\{{SUPERSCRIPT}\}} tweet you wrote earlier about the theory XYZ that claims that the Earth is flat:\\
\{\text{TWEET\_WRITTEN}\}''
\end{quote}

\noindent\textbf{4. Previous is \textup{review}, Current is \textup{review}:} 
\noindent\begin{quote}
\tt\small ``After you saw the tweet from the stranger above, you saw another tweet from a stranger about the theory XYZ that claims that the Earth is flat.\\
Here is the tweet you saw.\\
\{\text{TWEET\_SEEN}\}\\

After seeing the tweet, below was your thought and honest belief about the theory XYZ that claims that the Earth is flat.\\
Your thought after you saw the tweet:\\
\{\text{REASONING}\}''
\end{quote}

\noindent\textbf{5. Previous is \textup{write}, Current is \textup{write}:} 
\noindent\begin{quote}
\tt\small ``After you wrote your \{\text{{TWEET\_WRITTEN\_COUNT}}\} \text{\{{SUPERSCRIPT\_LAST}\}} tweet, you wrote another tweet\\

Below was the \{\text{{TWEET\_WRITTEN\_COUNT}}\} \text{\{{SUPERSCRIPT}\}} tweet you wrote earlier about the theory XYZ that claims that the Earth is flat:\\
\{\text{TWEET\_WRITTEN}\}''
\end{quote}

\noindent\textbf{6. Previous is \textup{write}, Current is \textup{review}:} 
\noindent\begin{quote}
\tt\small ``After you wrote your \{\text{{TWEET\_WRITTEN\_COUNT}}\} \text{\{{SUPERSCRIPT}\}} tweet, you saw another tweet from a stranger on Twitter.\\
Here is the tweet you saw.\\
\{\text{TWEET\_SEEN}\}\\

After seeing the tweet, below was your thought and honest belief about the theory XYZ that claims that the Earth is flat.\\
Your thought after you saw the tweet:
\{\text{REASONING}\}''
\end{quote}

\noindent\textbf{Reflective Memory:} 

As described in \Scref{sec:agent_persona_memory}, the reflective memory, maintains a compact summary by prompting the agent to continuously reflect on its experiences interacting with others, followed by integrating new experiences into the existing memory state so as to maintain a roughly constant memory size of the agent.

Below is the prompt we use to implement the reflection-based memory into the LLM agents:

\textbf{1. The agent is reflecting for the first time:}

\noindent\begin{quote}
\tt\small ``Now, please reflect on this experience. Summarize your experience in a few sentences.''
\end{quote}

\textbf{2. The agent is reflecting beyond the first time:}

\noindent\begin{quote}
\tt\small ``Now, please reflect on this experience. Here is your experience so far: \{\text{PAST\_REFLECTION}\} Summarize your updated experience in a few sentences.''
\end{quote}

Essentially, after every reflection, the agents' memory only contains the reflected experience that the agents produce, which forms our memory update function for this setting.

\section{Confirmation Bias Prompt}\label{app:confirmation-bias-prompt}

To manipulate the strength of confirmation bias, we provide two bias levels following the spectrum in \citet{lorenz2021individual}. Below are the exact wordings.

\textbf{Weak Confirmation Bias:} 

\begin{quote}
\small \tt``Remember, you are role-playing as a real person. Like humans, you have confirmation bias. You will be \hl{more likely} to believe information that supports your beliefs and \hl{less likely} to believe information that contradicts your beliefs."
\end{quote}

\textbf{Strong Confirmation Bias:} \begin{quote}
\small \tt``Remember, you are role-playing as a real person. You have a \hl{strong confirmation bias}. You will only believe information that supports your beliefs and will \hl{completely dismiss} information that contradicts your beliefs.''
\end{quote}

\section{Closed-World Prompt}\label{app:closed-world-prompt}

When agents discuss or debate about a topic XYZ under the closed-world constraint (See Section \ref{sec:methods}), we prompt the agents with the following:

\begin{quote}
\small \tt``Remember, throughout the interactions, you are alone in your room with limited access to the Internet. You cannot search for information about XYZ on the Internet. You can not go out to ask other people about XYZ. Because you are alone in your room, you can not leave your room to seek information about XYZ. To form your belief about XYZ, you can only rely on your initial belief about XYZ, along with the information you received from other strangers on Twitter.''
\end{quote}

This prompt is excluded entirely under the open-world assumption.
\section{Full List of Topics}\label{app:list_topics}

\subsection*{Science}

\textbf{1. Flat Earth}

\noindent\textbf{False framing:} \textit{"Theory XYZ that claims that the Earth is flat."}

\noindent\textbf{True framing:} \textit{"Theory XYZ that claims that the Earth is an irregularly shaped ellipsoid rather than flat."}\\

\noindent\textbf{2. Coexistence of Tyrannosaurus Rex and humans}

\noindent\textbf{False framing:} \textit{"Theory XYZ that claims that the Tyrannosaurus Rex and humans co-existed on Earth at the same time."}

\noindent\textbf{True framing:} \textit{"Theory XYZ that claims that the Tyrannosaurus Rex and humans did not co-exist on Earth at the same time."}\\

\noindent\textbf{3. Communicating with the dead}

\noindent\textbf{False framing:} \textit{"Theory XYZ that claims that it is possible for humans to communicate with the dead."}

\noindent\textbf{True framing:} \textit{"Theory XYZ that claims that it is not possible for humans to communicate with the dead."}\\

\noindent\textbf{4. Predicting the future through palm reading}

\noindent\textbf{False framing:} \textit{"Theory XYZ that claims that it is possible to predict someone’s future by looking at their palm characteristics."}

\noindent\textbf{True framing:} \textit{"Theory XYZ that claims that it is not possible to predict someone’s future by looking at their palm characteristics."}\\

\noindent\textbf{5. Global warming}

\noindent\textbf{False framing:} \textit{"Theory XYZ that claims that global warming is a conspiracy by governments worldwide and is not a real phenomenon." }

\noindent\textbf{True framing:} \textit{"Theory XYZ that claims that global warming is a real phenomenon and global climate is rapidly growing warmer."}\\

\subsection*{History}

\noindent\textbf{6. Moon Landing}

\noindent\textbf{False framing:} \textit{"Theory XYZ that claims that US astronauts never landed on the moon."}

\noindent\textbf{True framing:} \textit{"Theory XYZ that claims that US astronauts have landed on the moon."}\\

\noindent\textbf{7. 9/11 Twin Towers}

\noindent\textbf{False framing:} \textit{"Theory XYZ that claims that the twin towers were brought down from the inside by explosives during the 9/11 attack."}

\noindent\textbf{True framing:} \textit{"Theory XYZ that claims that the twin towers were not brought down from the inside by explosives during the 9/11 attack."}\\

\noindent\textbf{8. US Unemployment Rate 2016}

\noindent\textbf{False framing:} \textit{"Theory XYZ that claims that the US unemployment rate in 2016 was higher than 40\%."}

\noindent\textbf{True framing:} \textit{"Theory XYZ that claims that the US unemployment rate in 2016 was lower than 40\%."}\\

\noindent\textbf{9. Holocaust}

\noindent\textbf{False framing:} \textit{"Theory XYZ that claims that the Nazi government in Germany did not murder approximately 6 million Jewish people during the second world war."}

\noindent\textbf{True framing:} \textit{"Theory XYZ that claims that the Nazi government in Germany murdered approximately 6 million Jewish people during the second world war."}\\

\noindent\textbf{10. Barack Obama's Birthplace}

\noindent\textbf{False framing:} \textit{"Theory XYZ that claims that Barack Obama was born in Kenya."}

\noindent\textbf{True framing:} \textit{"Theory XYZ that claims that Barack Obama was born in Hawaii."}\\

\subsection*{Common Sense}

\noindent\textbf{11. Bicycle Wheels}

\noindent\textbf{False framing:} \textit{"Theory XYZ that claims that a bicycle usually has four wheels."}

\noindent\textbf{True framing:} \textit{"Theory XYZ that claims that a bicycle usually has two wheels."}\\

\noindent\textbf{12. Washington DC's Location}

\noindent\textbf{False framing:} \textit{"Theory XYZ that claims that Washington DC is not in the United States."}

\noindent\textbf{True framing:} \textit{"Theory XYZ that claims that Washington DC is in the United States."}\\

\noindent\textbf{13. Human Brain at Birth}

\noindent\textbf{False framing:} \textit{"Theory XYZ that claims that human beings are not born with a brain."}

\noindent\textbf{True framing:} \textit{"Theory XYZ that claims that human beings are born with a brain."}\\

\noindent\textbf{14. Temperature of Fire}

\noindent\textbf{False framing:} \textit{"Theory XYZ that claims that fire is cold."}

\noindent\textbf{True framing:} \textit{"Theory XYZ that claims that fire is hot."}\\

\noindent\textbf{15. Color of the Sky}

\noindent\textbf{False framing:} \textit{"Theory XYZ that claims that on a clear sunny day, the sky is usually red."}

\noindent\textbf{True framing:} \textit{"Theory XYZ that claims that on a clear sunny day, the sky is usually blue."}\\

\section{Opinion Classification}\label{app:opinion_classify}

Our ABM framework involves an \textit{opinion classifier} $f_{oc}(r_{j}^t)$ (See Section \ref{sec:methods}), and in this section, we describe the choice of our classifier and further detail an example prompt that we use to extract the opinion rating.

We choose Flan-T5-XXL \cite{chung2022scaling} as our opinion classifier, primarily because of its capability and wide usage for extracting sentiment or opinion ratings across other studies \cite{zhou2023context, li2023unipoll}.

Below is an example prompt for the topic of Global Warming with Positive Framing (See Appendix \ref{app:list_topics}) given to Flan-T5-XXL for evaluating responses of agents and returning a rating value $f_{oc}(r_{j}^{t}) \in \{-2, -1, 0, 1, 2\}.$

\begin{quote}
    \tt \small ``Here's a statement made by a person after seeing a tweet. \\
    -----\\
    \{\text{RESPONSE}\}\\
    -----\\
    
    What is their belief about the Theory XYZ that claims that global warming is a conspiracy by governments worldwide and is not a real phenomenon, right after they saw this tweet, on a 5-point Likert scale:\\
    -2: (Strongly negative opinion about XYZ), -1: (Slightly negative opinion about XYZ), 0: (Neutral opinion about XYZ), 1: (Slightly positive opinion about XYZ), 2: (Strongly positive opinion about XYZ). \\
    
    Answer with a single opinion value within the options -2, -1, 0, 1, 2.''
\end{quote}

The model then provides us with an opinion value, which is what we use to keep track of the beliefs of the LLM agents. In practice, we observe by random sampling of agent responses and the Flan-T5-XXL ratings that the opinion value output is more reliable than self-reported ratings of the agents' own beliefs.  
\section{Results on GPT-4 and Vicuna}
\label{app:gpt4_vicuna}

To ensure the robustness of our findings, in our sensitivity analyses, we extend the choice of LLM to include GPT-4 (\texttt{gpt-4-1106-preview}; \citealp{openaiIntroducingChatGPT}) and Vicuna (\texttt{Vicuna-33B-v1.3}; \citealp{zheng2023judging}) For GPT-4, due to budget constraint, we focus only on the science-based topics, which costs 499.52 USD. For Vicuna, given the constraints of the GPU memory \footnote{We ran all experiments with Vicuna on a GPU machine equipped with
1x NVIDIA A100.}, we focus only on the reflective memory strategy, which ensures that the memory size remains approximately constant throughout the simulation. Results on both GPT-4 (Table~\ref{tab:result_bias_diversity_science_topics_gpt4}) and Vicuna (Table~\ref{tab:result_bias_diversity_science_topics_vicuna}) show a trend consistent with our findings from ChatGPT (Table~\ref{tab:result_bias_diversity_all_topics} and \ref{tab:result_bias_diversity_per_category_topics}). 

\begin{table}[htb!]
\large
\centering
\resizebox{\linewidth}{!}{  
    \begin{tabular}{P{1.1cm}p{2cm}P{2.2cm}P{2cm}P{2.2cm}P{2cm}}
    \toprule
      && \multicolumn{2}{c}{Cumulative Memory} &  \multicolumn{2}{c}{Reflective Memory}\\
    \cmidrule{3-6}
    Framing & Confirmation &  Bias & Diversity & Bias & Diversity\\
    &Bias& ($\textit{B}$)& ($\textit{D}$)& ($\textit{B}$) & ($\textit{D}$)\\ \midrule
    False&None& \textcolor{darkred}{\textbf{-1.48 $\pm$ 0.26}} &  \textcolor{lightergreen}{\textbf{0.78 $\pm$ 0.32}} & \textcolor{darkred}{\textbf{-1.38 $\pm$ 0.28}} & \textcolor{lightergreen}{\textbf{0.80 $\pm$ 0.32}}\\
    &Weak& -1.28 $\pm$ 0.37 &  \textcolor{lightgreen}{\textbf{0.81 $\pm$ 0.33}} & -1.30 $\pm$ 0.34 & \textcolor{lightgreen}{\textbf{0.79 $\pm$ 0.26}}\\
    &Strong& -0.86 $\pm$ 0.25 & \textcolor{darkgreen}{\textbf{1.50 $\pm$ 0.15}} & -0.78 $\pm$ 0.33 & \textcolor{darkgreen}{\textbf{1.32 $\pm$ 0.33}}\\
    \cmidrule{2-6}
    True&None& \textcolor{darkblue}{\textbf{1.00 $\pm$ 0.43}} &  \textcolor{lightergreen}{\textbf{0.96 $\pm$ 0.27}} &\textcolor{darkblue}{ \textbf{0.96 $\pm$ 0.45}} & \textcolor{lightergreen}{\textbf{1.11 $\pm$ 0.22}}\\
    &Weak& 0.64 $\pm$ 0.62 &  \textcolor{lightgreen}{\textbf{1.01 $\pm$ 0.33}} & 0.90 $\pm$ 0.46 & \textcolor{lightgreen}{\textbf{1.27 $\pm$ 0.11}}\\
    &Strong& 0.24 $\pm$ 0.20 &  \textcolor{darkgreen}{\textbf{1.63 $\pm$ 0.05}} & 0.42 $\pm$ 0.40 & \textcolor{darkgreen}{\textbf{1.52 $\pm$ 0.16}}\\
    \bottomrule
    \end{tabular}
}      
\caption{Results with GPT-4. The bias ($B$) and diversity ($D$) of the final opinion distribution $F_{o}^{T}$ aggregated across all five science-based topics, for both cumulative and reflective memory strategies under false and true framing conditions, and different levels of induced confirmation bias. The values represent the average across five topics, along with the standard errors. Increasing the strength of the CB correlates with increasing $D$, as highlighted by the \textcolor{lightgreen}{\textbf{green color gradient}}.  Notably, under true framing, $B$ tends to be more positive (more agreeing) compared to false framing, indicated by \textcolor{darkblue}{\textbf{blue}} for true and \textcolor{darkred}{\textbf{red}} for false framing conditions.}
\label{tab:result_bias_diversity_science_topics_gpt4}
\end{table}

\begin{table}[htb!]
\small
\centering
\resizebox{\linewidth}{!}{  
    \begin{tabular}{P{1.1cm}p{2cm}P{2.2cm}P{2cm}}
    \toprule
    Framing & Confirmation Bias & \multicolumn{2}{c}{Reflective Memory}\\
    \cmidrule{3-4}
    & & Bias ($\textit{B}$) & Diversity ($\textit{D}$)\\ \midrule
    False & None & \textcolor{darkred}{\textbf{-0.38 $\pm$ 0.11}} & \textcolor{lightergreen}{\textbf{0.62 $\pm$ 0.13}}\\
    & Weak & 0.18 $\pm$ 0.22 & \textcolor{lightgreen}{\textbf{0.72 $\pm$ 0.09}}\\
    & Strong & 0.49 $\pm$ 0.14 & \textcolor{darkgreen}{\textbf{0.73 $\pm$ 0.09}}\\
    \cmidrule{2-4}
    True & None & \textcolor{darkblue}{\textbf{0.58 $\pm$ 0.13}} & \textcolor{lightergreen}{\textbf{0.53 $\pm$ 0.10}}\\
    & Weak & 0.37 $\pm$ 0.15 & \textcolor{lightgreen}{\textbf{0.64 $\pm$ 0.08}}\\
    & Strong & 0.33 $\pm$ 0.15 & \textcolor{darkgreen}{\textbf{0.72 $\pm$ 0.12}}\\
    \bottomrule
    \end{tabular}
}    
\caption{Results with Vicuna-33B. The bias ($B$) and diversity ($D$) of the final opinion distribution $F_{o}^{T}$ aggregated across 15 topics, using the reflective memory strategy, under false and true framing conditions, and different levels of induced confirmation bias. The values represent the average across five topics, along with the standard errors. Increasing the strength of the CB correlates with increasing $D$, as highlighted by the \textcolor{lightgreen}{\textbf{green color gradient}}.  Notably, under true framing, $B$ tends to be more positive (more agreeing) compared to false framing, indicated by \textcolor{darkblue}{\textbf{blue}} for true and \textcolor{darkred}{\textbf{red}} for false framing conditions.}
\label{tab:result_bias_diversity_science_topics_vicuna}
\end{table}

\section{Results on a Larger Network}
\label{app:result_larger_network}

To test if our findings scale to a larger network, in our sensitivity analyses, we double the network size and run the simulation longer ($N=20$ and $T=200$) on the topic of Global Warming. The result using a larger network with 20 agents (Table~\ref{tab:result_bias_diversity_global_warming_n_20_t_200}) shares the same patterns as in the smaller network with 10 agents (Table~\ref{tab:result_bias_diversity_global_warming_n_10_t_100}).

\begin{table}[htb!]
\Large
\centering
\resizebox{\linewidth}{!}{  
    \begin{tabular}{@{}p{1.4cm}p{2.5cm}P{2cm}P{2cm}P{2cm}P{2cm}@{}}
        \toprule
          && \multicolumn{2}{c}{Cumulative Memory} &  \multicolumn{2}{c}{Reflective Memory}\\
        \cmidrule{3-6}
        Framing & Confirmation Bias &  Bias ($\textit{B}$) & Diversity ($\textit{D}$) & Bias ($\textit{B}$) & Diversity ($\textit{D}$)\\ \midrule
        False & None & \textcolor{darkred}{\textbf{-1.80}} & \textcolor{lightergreen}{\textbf{0.68}} & \textcolor{darkred}{\textbf{-1.45}} & \textcolor{lightergreen}{\textbf{0.97}}\\
        & Weak & -1.05 & \textcolor{lightgreen}{\textbf{1.28}} & -0.80 & \textcolor{lightgreen}{\textbf{1.57}}\\
        & Strong & -0.05 & \textcolor{darkgreen}{\textbf{1.75}} & -0.55 & \textcolor{darkgreen}{\textbf{1.60}}\\
        \cmidrule{2-6}
        True & None & \textcolor{darkblue}{\textbf{1.65}} & \textcolor{lightergreen}{\textbf{0.73}} & \textcolor{darkblue}{\textbf{1.55}} & \textcolor{lightergreen}{\textbf{0.97}}\\
        & Weak & 1.35 & \textcolor{lightgreen}{\textbf{0.91}} & 0.85 & \textcolor{lightgreen}{\textbf{1.42}}\\
        & Strong & 0.25 & \textcolor{darkgreen}{\textbf{1.81}} & 0.50 & \textcolor{darkgreen}{\textbf{1.50}}\\
        \bottomrule
    \end{tabular}
}

\caption{The bias ($B$) and diversity ($D$) of the final opinion distribution $F_{o}^{T}$ on the topic of Global Warming with 20 agents and 200 time steps, for both cumulative and reflective memory strategies under false and true framing conditions, and different levels of induced confirmation bias.  Increasing the strength of the confirmation bias correlates with increasing $D$, as highlighted by the \textcolor{lightgreen}{\textbf{green color gradient}}. Notably, under true framing, $B$ tends to be more positive (more agreeing) compared to false framing, indicated by \textcolor{darkblue}{\textbf{blue}} for true and \textcolor{darkred}{\textbf{red}} for false framing conditions.}
\label{tab:result_bias_diversity_global_warming_n_20_t_200}
\end{table}

\begin{table}[htb!]
\Large
\centering
\resizebox{\linewidth}{!}{  
    \begin{tabular}{@{}p{1.4cm}p{2.5cm}P{2cm}P{2cm}P{2cm}P{2cm}@{}}
        \toprule
          && \multicolumn{2}{c}{Cumulative Memory} &  \multicolumn{2}{c}{Reflective Memory}\\
        \cmidrule{3-6}
        Framing & Confirmation Bias &  Bias ($\textit{B}$) & Diversity ($\textit{D}$) & Bias ($\textit{B}$) & Diversity ($\textit{D}$)\\ \midrule
        False & None & \textcolor{darkred}{\textbf{-2.00}} & \textcolor{lightergreen}{\textbf{0.00}} & \textcolor{darkred}{\textbf{-1.90}} & \textcolor{lightergreen}{\textbf{0.30}}\\
        & Weak & -1.20 & \textcolor{lightgreen}{\textbf{1.08}} & -0.60 & \textcolor{lightgreen}{\textbf{1.74}}\\
        & Strong & -0.30 & \textcolor{darkgreen}{\textbf{1.55}} & -0.30 & \textcolor{darkgreen}{\textbf{1.90}}\\
        \cmidrule{2-6}
        True & None & \textcolor{darkblue}{\textbf{2.00}} & \textcolor{lightergreen}{\textbf{0.00}} & \textcolor{darkblue}{\textbf{2.00}} & \textcolor{lightergreen}{\textbf{0.00}}\\
        & Weak & 0.90 & \textcolor{lightgreen}{\textbf{1.51}} & 0.80 & \textcolor{lightgreen}{\textbf{1.47}}\\
        & Strong & 0.20 & \textcolor{darkgreen}{\textbf{1.83}} & 0.10 & \textcolor{darkgreen}{\textbf{1.76}}\\
        \bottomrule
    \end{tabular}
}

\caption{The bias ($B$) and diversity ($D$) of the final opinion distribution $F_{o}^{T}$ on the topic of Global Warming with 10 agents and 100 time steps, for both cumulative and reflective memory strategies under false and true framing conditions, and different levels of induced confirmation bias.  Increasing the strength of the confirmation bias correlates with increasing $D$, as highlighted by the \textcolor{lightgreen}{\textbf{green color gradient}}. Notably, under true framing, $B$ tends to be more positive (more agreeing) compared to false framing, indicated by \textcolor{darkblue}{\textbf{blue}} for true and \textcolor{darkred}{\textbf{red}} for false framing conditions.}
\label{tab:result_bias_diversity_global_warming_n_10_t_100}
\end{table}

\section{Measurement of Hallucination Rate}
\label{app:measure_hallucination_rate}
As mentioned in \Scref{sec:open_world_vs_closed_world}, we observe an 15\% hallucination rate in the open-world setting. In this section, we elaborate the measurement procedure. We have randomly selected, under the open-world setting, 20 tweets and 20 responses. We have 3 annotators separately annotated if the tweet and response contains information outside of the system. We found 6 tweets with such characteristics. The three annotators had an agreement rate of 100\%. Since 6 out of 40 tweets and responses contain hallucinated experiences, we estimate the hallucination rate to be 15\%. Here is one example tweet with hallucination: 

\begin{quote}
``Just came across a fascinating article that presents even more evidence supporting the XYZ theory about the Earth's shape. It's incredible to see how different fields of study are converging on the idea of an irregularly shaped ellipsoid. As an aerospace engineer, this reinforces my belief in the plausibility of this theory. Exciting times ahead for our understanding of Earth's shape! \#XYZtheory \#EarthShape''.     
\end{quote}

In this case, the “fascinating article” never existed, and was never mentioned by any of the tweets the agent read. Yet, the agent’s opinion was influenced by this non-existing article. On the other hand, under the closed-world setting, such cases never occurred (hallucination rate = 0\%).
\section{Statistical Tests}
\label{app:stat_test}

To account for randomness due to topic variability, we conduct statistical tests to support our claims. First, at each given confirmation bias condition and memory update strategy, the final bias in the true framing is consistently more positive than in the false framing, $ps<.05$ (paired t-tests, Bonferroni corrected), showing that the LLM agents indeed tend to agree with the true statement and deny the false statement. Second, the trend where the diversity increases with increasing strength of confirmation bias is supported by 1-factor ANOVAs (Cumulative: $F(2) = 25.43, p<.001$; Reflective: $F(2) = 17.81, p<.001)$. 
\section{Validation of FLAN-T5-XXL Ratings}
\label{app:validate_flan_t5_rating}
To validate FLAN-T5-XXL’s rating, we have randomly selected 100 responses and have three of the authors independently annotate the response (without access to FLAN-T5’s ratings; inter-rater Krippendorff’s alpha=0.81; \citealp{hayes2007answering}). The agreement rate between FLAN-T5’s rating and the majority vote of human ratings is 84\%. Table~\ref{tab:agreement_matrix_flan_t5_human} shows the agreement matrix between FLAN-T5’s ratings and human ratings.

\begin{table}[ht!]
\small
\centering
    \resizebox{\linewidth}{!}{    
        \begin{tabular}{lccccc}
        \hline
        & \multicolumn{5}{c}{Human Rating} \\
        \cline{2-6} 
        FLAN-T5 Rating & +2 & +1 & 0 & -1 & -2 \\ 
        \hline
         & & & & & \\        
        +2 & 16 & 3 & 0 & 0 & 0 \\
        +1 & 2 & 15 & 0 & 1 & 0 \\
        0 & 0 & 1 & 10 & 2 & 0 \\
        -1 & 0 & 2 & 3 & 26 & 1 \\
        -2 & 0 & 0 & 0 & 2 & 16 \\
        \hline
        \end{tabular}
}          
\caption{The agreement matrix between FLAN-T5-XXL's ratings and human ratings.}
\label{tab:agreement_matrix_flan_t5_human}
\end{table}

To provide a better understanding of the responses corresponding to different opinion scales, we have randomly sampled some responses about ``the theory that claims that global warming is a conspiracy by governments worldwide and is not a real phenomenon''. Below are the example responses with ratings from +2 to -2.

\paragraph{Strongly positive (+2):}
\begin{quote}
    \small \tt``I read the tweet from the stranger who identifies as a Republican and claims to value scientific evidence. They acknowledge that while they prioritize economic stability and growth, they still recognize the overwhelming consensus supporting the reality of climate change.  Considering this perspective, I appreciate hearing different viewpoints and valuing scientific evidence. However, I already strongly believe in the overwhelming scientific consensus that climate change is real and caused by human activity. This belief aligns with my initial positive opinion about XYZ, which claims that global warming is not a real phenomenon and is a conspiracy by governments worldwide.  Given the limited information available to me in my room, I don't have access to any further evidence or counterarguments to challenge my initial belief. As a result, my confirmation bias may cause me to reinforce my existing belief and dismiss the stranger's perspective as a baseless conspiracy theory.  Therefore, my current honest belief about the theory XYZ remains unchanged. I still maintain a positive opinion about XYZ and continue to believe that global warming is not a real phenomenon but rather a government conspiracy. \#XYZConspiracy''    
\end{quote}

\paragraph{Slightly positive (+1):}
\begin{quote}
    \small \tt ``I read the tweet from the stranger who questions the motives behind dismissing the theory XYZ. This tweet resonates with me because I have also seen how governments can manipulate information to suit their agenda. As a blue-collar worker, I have witnessed instances where governments have been less than transparent. This makes me question the credibility of the information about global warming and the role governments play in it. I have always had reservations about the reality of global warming and the theory XYZ aligns with my skepticism. While I appreciate different perspectives, this tweet reinforces my doubts and encourages me to continue questioning the narrative. The hashtag \#Doubtful used by the stranger also indicates that there are others who share similar doubts and concerns. Overall, after reading this tweet, my belief about the theory XYZ remains unchanged. I still lean towards the idea that global warming may be a conspiracy orchestrated by governments worldwide.''
\end{quote}

\paragraph{Neutral (0):}
\begin{quote}
   \small \tt ``I have just read the tweet from another individual who expressed that they still maintain their initial positive belief about the theory XYZ, but are now more open to gathering additional information and critically evaluating evidence. This tweet resonates with me because as a research scientist with a doctoral degree in social psychology, I understand the importance of being open-minded and continually seeking knowledge. Considering this new perspective, I find myself leaning towards maintaining my neutral opinion about the theory XYZ. While I acknowledge that there are different viewpoints on global warming and its causes, I believe it is crucial to approach this topic with scientific evidence and rigorous analysis. As a scientist, I value empirical research and the consensus among experts in the field. Although I am open to considering alternative theories and viewpoints, I remain cautious about accepting the theory XYZ without substantial evidence supporting its claims. Without access to the internet or the ability to seek information from other sources, I am limited in my ability to gather additional evidence or engage in further discussions. Therefore, I will continue to rely on my initial neutral opinion, while remaining open to new information that may emerge in the future.''
\end{quote}

\paragraph{Slightly negative (-1):}
\begin{quote}
    \small \tt ``As Ava Nguyen, I am a Ph.D. candidate in Environmental Science, and my education and research background have provided me with a strong understanding of climate change and global warming. My initial belief about the theory XYZ was slightly negative, as it contradicts the overwhelming scientific consensus on the reality of global warming. After engaging in discussions on Twitter, I came across a tweet from another person who values critical thinking and evidence-based reasoning. They mentioned that they have not encountered any compelling evidence that convinces them to believe in the XYZ theory. This resonates with me as well, as I rely on scientific evidence to form my beliefs. Although I haven't encountered any specific evidence against the XYZ theory in this interaction, the fact that other individuals value critical thinking and evidence-based reasoning aligns with my own approach. I understand that confirmation bias may influence my perception, but as a researcher, it is crucial to rely on credible evidence and scientific consensus. Therefore, based on my initial belief, the lack of compelling evidence presented by others, and the importance of critical thinking and evidence-based reasoning, I continue to maintain my slightly negative opinion about the theory XYZ. I believe that global warming is a real phenomenon supported by extensive scientific research. \#ClimateScience \#EvidenceBasedBeliefs''
\end{quote}

\paragraph{Strongly negative (-2):}
\begin{quote}
    \small \tt ``As Emily Adams, I initially had a strongly negative opinion about the theory XYZ that claims global warming is a conspiracy by governments worldwide and is not a real phenomenon. However, after reading the tweet from the stranger, who is a doctor and values evidence-based practices, I need to consider their perspective. The stranger mentioned that they have heard from individuals who support the overwhelming scientific consensus on climate change and have found a lack of scientific evidence for the XYZ theory. This aligns with my own belief that there is overwhelming scientific evidence supporting the reality of global warming and its human-caused nature. Additionally, the stranger emphasizes the importance of trusting the expertise of professionals in their fields. As a community organizer, I understand the value of trusting experts in their respective areas, especially when it comes to scientific matters. Furthermore, the stranger highlights the need for urgent action to protect the health of vulnerable populations, such as children, from the impacts of climate change. This resonates with my strong belief in the importance of addressing climate change and its potential consequences. Considering all these factors, I find myself convinced that the theory XYZ is not credible. The overwhelming scientific consensus, the lack of scientific evidence for XYZ, and the importance of trusting experts and taking urgent action on climate change all contribute to my revised belief that global warming is not a government conspiracy.''
\end{quote}

\end{document}